\documentclass[fleqn,usenatbib]{mnras}

\usepackage{newtxtext,newtxmath}
\usepackage[T1]{fontenc}
\usepackage{ae,aecompl}

\usepackage{graphicx}	% Including figure files
\usepackage{amsmath}	% Advanced maths commands
\usepackage{subcaption}
\usepackage{pdflscape}

\usepackage{scalerel,tikz}
\usetikzlibrary{svg.path}
\definecolor{orcidlogocol}{HTML}{A6CE39}
\tikzset{orcidlogo/.pic={
 \fill[orcidlogocol] svg{M256,128c0,70.7-57.3,128-128,128C57.3,256,0,198.7,0,128C0,57.3,57.3,0,128,0C198.7,0,256,57.3,256,128z};
 \fill[white] svg{M86.3,186.2H70.9V79.1h15.4v48.4V186.2z}
 svg{M108.9,79.1h41.6c39.6,0,57,28.3,57,53.6c0,27.5-21.5,53.6-56.8,53.6h-41.8V79.1z M124.3,172.4h24.5c34.9,0,42.9-26.5,42.9-39.7c0-21.5-13.7-39.7-43.7-39.7h-23.7V172.4z}
 svg{M88.7,56.8c0,5.5-4.5,10.1-10.1,10.1c-5.6,0-10.1-4.6-10.1-10.1c0-5.6,4.5-10.1,10.1-10.1C84.2,46.7,88.7,51.3,88.7,56.8z};
}}
\newcommand\orcidicon[1]{\href{https://orcid.org/#1}{\mbox{\scalerel*{
\begin{tikzpicture}[yscale=-1,transform shape]
\pic{orcidlogo};
\end{tikzpicture}
}{|}}}}

\newcommand{\red}{\textcolor{red}}

\title[Observational window effects on multi-object RM]{Observational window effects on multi-object Reverberation Mapping}

\author[Malik et al.]{Umang Malik$^{\orcidicon{0000-0002-0036-1696}\,1}$\thanks{umang.malik@anu.edu.au},
Rob Sharp$^{\orcidicon{0000-0003-4877-7866}\,1}$\thanks{rob.sharp@anu.edu.au}, 
Paul Martini$^{\orcidicon{0000-0002-4279-4182}\,2,3,4}$,
Tamara M.~Davis$^{\orcidicon{0000-0002-4213-8783}\,5}$,
Brad E. Tucker$^{\orcidicon{0000-0002-4283-5159}\,1,6,7}$,
\newauthor
Zhefu Yu$^{\orcidicon{0000-0003-0644-9282}\,2}$,
Andrew Penton$^{\orcidicon{0000-0002-9999-7757}\,5}$,
Geraint F. Lewis$^{\orcidicon{0000-0003-3081-9319}\,8}$ and
Josh Calcino$^{\orcidicon{0000-0001-7764-3627}\,5}$
\\
$^{1}$Research School of Astronomy and Astrophysics, Australian National University, Canberra, ACT 2611, Australia\\
$^{2}$Department of Astronomy, The Ohio State University, Columbus, Ohio 43210, USA\\
$^{3}$Center of Cosmology and Astro-Particle Physics, The Ohio State University, Columbus, Ohio 43210, USA\\
$^{4}$Radcliffe Institute for Advanced Study, Harvard University, Cambridge, MA 02138, USA\\
$^{5}$School of Mathematics and Physics, The University of Queensland,  St Lucia, QLD 4101, Australia\\
$^{6}$National Centre for the Public Awareness of Science, Australian National University, Canberra, ACT 2601, Australia\\
$^{7}$The ARC Centre of Excellence for All-Sky Astrophysics in 3 Dimension (ASTRO 3D), Australia\\
$^{8}$Sydney Institute for Astronomy, School of Physics, A28, The University of Sydney, NSW 2006, Australia
}

\date{Accepted XXX. Received YYY; in original form \today}

\pubyear{2022}

\begin{document}
\label{firstpage}
\pagerange{\pageref{firstpage}--\pageref{lastpage}}
\maketitle

% Abstract of the paper
\begin{abstract}
Contemporary reverberation mapping campaigns are employing wide-area photometric data and high-multiplex spectroscopy to efficiently monitor hundreds of active galactic nuclei (AGN). However, the interaction of the window function(s) imposed by the observation cadence with the reverberation lag and AGN variability time scales (intrinsic to each source over a range of luminosities) impact our ability to recover these fundamental physical properties. Time dilation effects due to the sample source redshift distribution introduces added complexity. We present comprehensive analysis of the implications of observational cadence, seasonal gaps and campaign baseline duration (\textit{i.e.,} the survey window function) for reverberation lag recovery. We find the presence of a significant seasonal gap dominates the efficacy of any given campaign strategy for lag recovery across the parameter space, particularly for those sources with observed-frame lags above 100 days. Using the OzDES survey as a baseline, we consider the implications of this analysis for the 4MOST/TiDES campaign providing concurrent follow-up of the LSST deep-drilling fields, as well as upcoming programs. We conclude that the success of such surveys will be critically limited by the seasonal visibility of some potential field choices, but show significant improvement from extending the baseline. Optimising the sample selection to fit the window function will improve survey efficacy.
% This is a simple template for authors to write new MNRAS papers.
% The abstract should briefly describe the aims, methods, and main results of the paper.
% It should be a single paragraph not more than 250 words (200 words for Letters).
% No references should appear in the abstract.
\end{abstract}

% Select between one and six entries from the list of approved keywords.
% Don't make up new ones.

%%%%%%%%%%%%%%%%%%%%%%%%%%%%%%%%%%%%%%%%%%%%%%%%%%

%%%%%%%%%%%%%%%%% BODY OF PAPER %%%%%%%%%%%%%%%%%%

\section{Introduction}

The innermost regions of active galactic nuclei (AGN) provide a crucial window for the understanding of supermassive black hole (SMBH) physics and galaxy evolution. Within the local Universe, this zone can be resolved to within the gravitational sphere of influence through the use of the highest angular resolution instruments \citep[e.g.,][]{Gebhardt2000,Greene2010,Gebhardt2011,Kuo2011,EHT2019}. Beyond the local Universe the achievable angular resolution becomes inadequate to resolve the environments of even the most massive black hole systems. We must turn to indirect techniques to probe the structure of such systems.

Reverberation Mapping (RM) provides a powerful opportunity to probe the structure of even the most distant of these compact systems. By resolving the innermost regions of AGN in the time-domain, the variable continuum emission from an accretion disk triggers a {\it reverberation} response in the surrounding environment and this signal can be inverted to reveal otherwise unresolvable physical detail. The broad line region (BLR) produces an emission-line spectrum in which the flux of the emission lines reverberates in response to the variability of the ionising continuum emission \citep{Blandford1982, Peterson2004}. The delay $\tau$, due to light travel time between the continuum emission from the compact core and the emission from the BLR, allows the radius of the BLR ($R_{\textrm{BLR}}\sim c \tau$) to be inferred. The velocity dispersion of the BLR ($\Delta V$) can be estimated from the width of the broadened emission lines. The mass of the central black hole ($M_{\textrm{BH}}$) can then be measured using the virial theorem:
\begin{equation}\label{virial}
    M_{\textrm{BH}} = f\frac{R_{\textrm{BLR}} \Delta V^2}{G},
\end{equation}
where $f$ is the virial coefficient; a dimensionless scale factor that accounts for the geometry, orientation, and kinematics of the BLR \citep{Woo2015}.

This technique has been used to measure the masses of over 100 SMBHs, establishing RM as the primary method of SMBH mass measurement at $z > 0.1$ at present. The tight correlation found between the AGN luminosity and the radius of the BLR \citep[$R-L$ relation; e.g.,][]{Bentz2009} has facilitated the use of indirect virial BH mass estimators using single-epoch spectroscopy \citep[e.g.,][]{Shen2011}. \citet{Watson2011} proposed using the $R-L$ relation to standardise AGN for use in cosmology. Although RM is a powerful technique with many applications, its full potential is yet to be reached. Few measurements have been made for the Mg~\textsc{ii} and C~\textsc{iv} emission lines, preventing the $R-L$ relations for these lines from being better constrained. This has also meant that the subset of AGN for which most RM measurements have been made are limited to local AGN, which are not representative of the general AGN population \citep{Shen2008,Shen2013}. It is necessary to probe the full range of AGN luminosities to improve virial BH mass estimates for the broader population.

Time-domain photometric and spectroscopic monitoring is required to successfully conduct RM of the BLR, since the technique relies on tracing the stochastic variability of AGN. To mitigate the bias towards local AGN, recent ‘industrial-scale’ RM campaigns have probed new regions of the AGN luminosity-redshift parameter space, with a particular focus at high-redshifts. This includes the Australian Dark Energy Survey (OzDES) RM Program \citep{King2015} and Sloan Digital Sky Survey RM Project \citep[SDSS-RM,][]{Shen2015}, which began in 2013 and 2014, respectively. Developments with multi-object spectroscopy have enabled these surveys to monitor large AGN samples. The success of lag recovery is highly dependent on the observational window, therefore the cadence, season length and baseline must be optimised to achieve the goals of the survey.

The observational window function (both photometric and spectroscopic) of any observing campaign imposes structure on the light curve sampling for a RM survey. Long baseline time-series photometry is readily obtainable through contemporary wide-field surveys (e.g., Dark Energy Survey, LSST). Obtaining concurrent spectroscopy is more challenging, due to the large multiplex of a wide field-of-view required. Facility choice is furthered restricted by the need for a large telescope aperture, essential to achieve sufficient signal-to-noise in the spectroscopic data for practical integration times. Improved facility access, and broad appeal across a range of survey applications, drives many surveys to equatorial fields to facilitate access to the survey program from both hemispheres. Observing equatorial fields immediately restricts field visibility and imposes a significant seasonal gap on the data.

The RM technique relies on recovering the lag between observed continuum and emission-line light curves using cross-correlation, which is highly dependent on how the light curves are sampled. The effect of the observational window on lag recovery has been studied in a limited capacity. \citet{White1994} investigated the minimum required sampling interval (\textit{i.e.,} cadence) for accurate recovery, and \citet{Welsh1999} identified improvement with extending the duration of light curves. However both these studies were limited to studying the recovery for local AGN with lags less than a month long, on an individual source basis. For contemporary large-scale surveys, \citet{King2015} and \citet{Shen2015} conducted simulations of the OzDES and SDSS-RM surveys, respectively. The simulations tested some program alterations and extensions, which highlighted improvements with extending the season length and baseline, but this was not used to alter the implemented survey design. Recent RM results from these programs \citep{Grier2017,Hoormann2019,Grier2019,Homayouni2020,Yu2021,Penton2022} show how the observational window presents challenges for recovery of these high-$z$ AGN lags, such as aliasing due to seasonal gaps. In addition, lag recovery depends on the signal-to-noise of the flux measurements and observed variability of the AGN. After combining these factors we see how complex lag recovery can be, and the need for a comprehensive study of how it can be improved. 

There are factors in the success of RM which can be controlled, and those that can not. It is possible that the observed variability of a source is simply unfavourable (e.g., monotonic behaviour, \citealt{Kaspi2007}), or that some sources do not reverberate ideally (e.g., BLR `holiday', \citealt{Goad2016,Dehghanian2019}). As more RM surveys are completed, a better understanding of such phenomena can be achieved. The signal-to-noise of spectroscopic measurements, particularly at high redshifts, also presents a challenge. Presently RM surveys must focus on the factors that can be controlled: the survey window. In this work, we present a detailed investigation of each component of the global window function which describes the survey sampling. In Section \ref{sec:modelWF} we describe the AGN light curve model, window functions, and lag recovery methods. Section \ref{sec:results} presents simulations designed to probe the 3-dimensional parameter space of cadence, season length and baseline to see how the accuracy and precision of the lag recovery changes. We then extend this to a subsample of simulated AGN with a range of expected lags before conducting high-resolution simulations for a grid of AGN probing a wide region of redshift-luminosity parameter space. We simulate the efficacy of a future planned `industrial-scale' RM survey and potential extensions that could improve its success. We discuss the implications and caveats of our work in Section \ref{sec:discussion} and summarise our key results in Section \ref{sec:summary}.

\section{Modelling observational window functions} \label{sec:modelWF}

The DES deep survey targeted legacy supernovae fields across a wide range in declination, with all regions followed up with multi-object spectroscopy conducted by OzDES. We model the resulting observational window function for RM using this contemporary survey as an initial baseline. The OzDES window function is largely representative of the window functions of current and planned large-scale RM surveys \citep[e.g.,][]{Shen2015,Kollmeier2017,Brandt2018,Swann2019,MSE2019}. The custom window functions we apply probe the window function parameter space by modelling surveys that could be performed by contemporary facilities. To conduct our simulations, we created mock AGN light curves to which we could apply our custom window functions. In this section we introduce the OzDES survey, used as our base model survey, and describe how we create mock light curves. We then explain the custom window functions, and the lag recovery methods used for this work.

\subsection{Australian Dark Energy Survey Reverberation Mapping Program}\label{sec:ozdes}

As part of the broader Dark Energy Survey (DES), the DES supernovae (SNe) program repeatedly observed 10 deep fields, covering 27 deg$^2$ of the sky, in the $g,r,i$ and $z$ filters, with the Dark Energy Camera (DECam) on the 4m Blanco telescope at Cerro Tololo Inter-American Observatory (CTIO) \citep{DES2016}. These fields comprise of the ELIAS, XMM-Large Scale Structure, Chandra deep-field South, and SDSS Stripe 82 fields. The deep fields were observed over a 5-6 month campaign season (from September to January) each year, from September 2013 to January 2019. The photometric observations were obtained, on average, every 6 days during each season. This cadence, designed principally for detection of transient events such as type-Ia SNe, provides continuum light curves for AGN reverberation mapping. Wide-field multi-object spectroscopy for the SNe program was provided by the OzDES program. OzDES performed the spectroscopic follow-up of the deep fields with the 2dF multi-object fibre positioning system and the AAOmega spectrograph \citep{Sharp2006} on the 3.9m Anglo-Australian Telescope (AAT), with a monthly cadence over the same 6-year baseline \citep{Yuan2015,Childress2017,Lidman2020}. The primary scientific goals of OzDES were to follow-up supernovae discovered by DES, by using their spectra to determine host-galaxy redshifts and for classification, and to conduct RM of AGN. 

\begin{figure}
	\centering
	\includegraphics[scale=0.39]{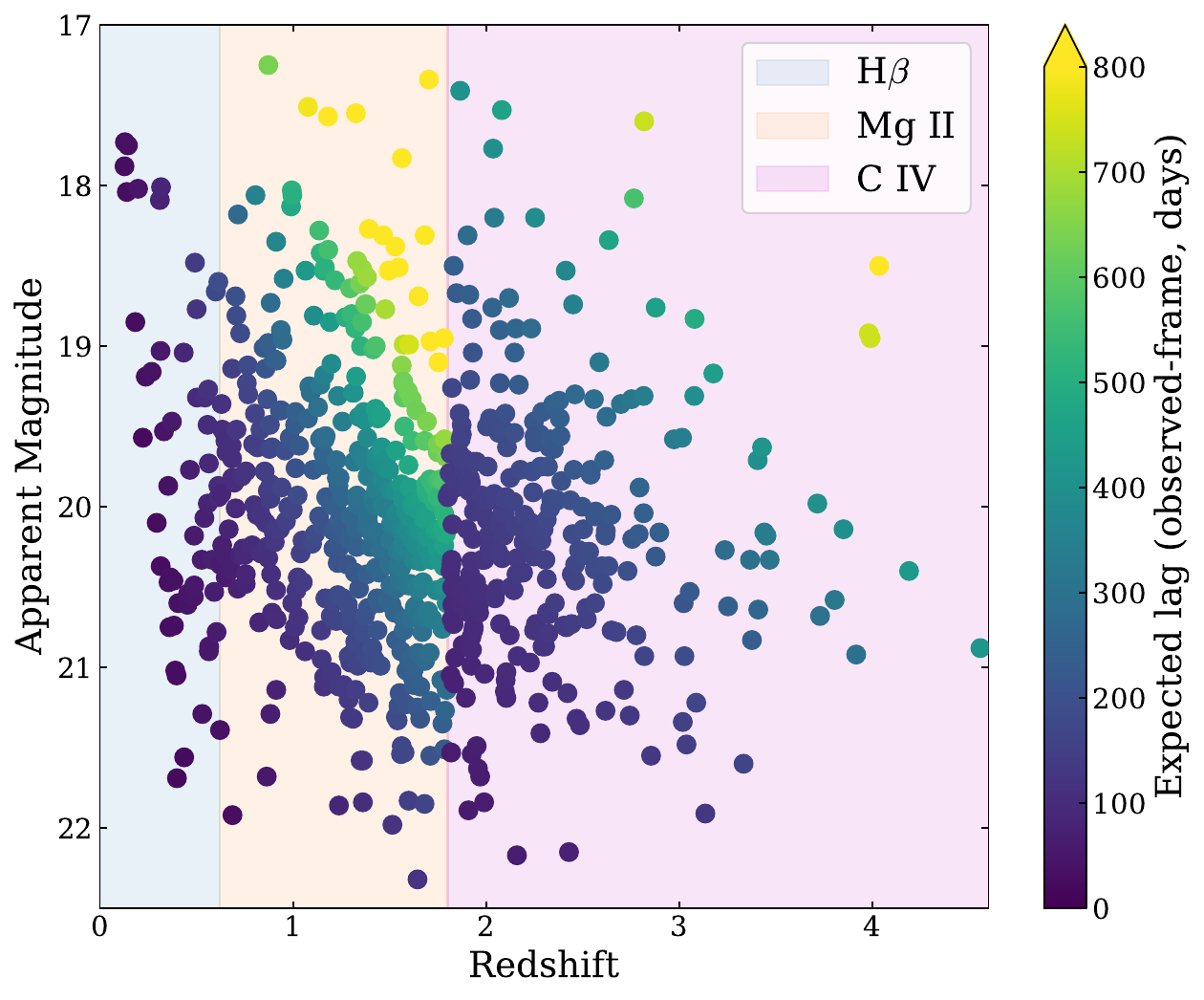}
	\caption{Distribution of redshifts and $r$ band apparent AB magnitudes for the 771 AGN monitored by OzDES for RM, and their expected observed-frame lags from published $R-L$ relations. The relations from \citet{Bentz2013}, \citet{Trakhtenbrot2012} and \citet{Hoormann2019} were used for the H$\beta$ (0$<z<$0.62), Mg\,\textsc{ii} (0.62$<z<$1.78) and C\,\textsc{iv} ($z>$1.78) emission lines, respectively.} 
    \label{fig:ozdes_AGN}
\end{figure}

As part of the OzDES RM program, 771 AGN have been monitored, at redshifts 0.12 $<$ $z$ $<$ 4.5 with apparent magnitudes 17.2 $<$ $r_{\rm{AB}}$ $<$ 22.3. The distributions of the redshifts and magnitudes of these targets are shown in \autoref{fig:ozdes_AGN}, which also shows the expected observed-frame lags for our sample. The sample was screened for sources that fell within DECam chip gaps, or with emission lines compromised by the transition between the red and blue arms of AAOmega \citep[see][for more details]{King2015,Tie2017}. The SDSS-RM sample follows a similar distribution \citep{Shen2015}. 

We continued photometric observations with DECam in 2019, and spectroscopic observations with the AAT in 2020. Due to COVID-19 disruptions closing the observatory in 2020, we continued the photometric observations with DECam in 2021. This will partially bridge the gap to the Legacy Survey of Space and Time (LSST) conducted with the Vera Rubin Observatory, which will also be observing the ELIAS, XMM-LSS and Chandra DFS, as three of its four deep-drilling fields. This will also provide continuity to the spectroscopic follow-up of the LSST deep-drilling fields by SDSS-V \citep{Kollmeier2017} and the Time-Domain Extragalactic Survey (TiDES; \citealt{Swann2019}), which we discuss in \S\ref{sec:para_space}.

\subsection{Light curves}

The damped random walk (DRW) model developed by \citet{Kelly2009} was used to create simulated continuum light curves. It has been shown to provide an adequate representation of the intrinsically stochastic behaviour exhibited by AGN, as seen in their continuum light curves \citep{Kozlowski2010, MacLeod2010}. The DRW model has been widely used for RM simulations \citep{Shen2015, King2015, Li2019, Penton2022}. Although AGN variability has been observed to deviate from the DRW on short time-scales \citep[][]{Mushotzky2011,Kasliwal2015,Smith2018}, this has been found to not affect lag measurements \citep{Yu2020}.

The procedure we used to generate the simulated light curves is detailed in \citet{Penton2022}. The required inputs are the source redshift and apparent magnitude, from which each of the model parameters are derived. For each simulated source, the expected rest-frame lag was estimated using the $R-L$ relations for the H$\beta$ (0$<z<$0.62) \citep{Bentz2013}, Mg\,\textsc{ii} (0.62$<z<$1.78) \citep{Trakhtenbrot2012} and C\,\textsc{iv} ($z>$1.78) \citep{Hoormann2019} emission lines, and multiplied by $(1+z)$ to obtain the input observed-frame lag, $\tau_{\rm input}$. Unless otherwise stated, the light curves, input lags and timescales we refer to are in the observed frame.   

The simulated AGN light curves were generated with daily cadence of the photometric and spectroscopic light curves, then appropriately down-sampled by the window function. The photometric and spectroscopic fluxes have 3\% and 10\% observational uncertainties applied, respectively, approximately representative of the DES and OzDES survey data \citep[][]{DES2016,Lidman2020}.

\subsection{Window functions}

The dominant observational effect on lag recovery is expected to be the survey window function. For our simulations we apply custom window functions to model observed light curves. We apply a custom window function that samples the original simulated light curves with a desired baseline, season length, and photometric or spectroscopic cadence (for the continuum and emission-line light curves, respectively). To be representative of the uneven sampling of surveys, we permit a scatter range for the cadence (e.g., 14$\,\pm\,$4 days), drawing the dates from a uniform distribution.

\subsection{Lag Recovery}

\begin{figure*}
	\centering
	\includegraphics[width=0.87\textwidth]{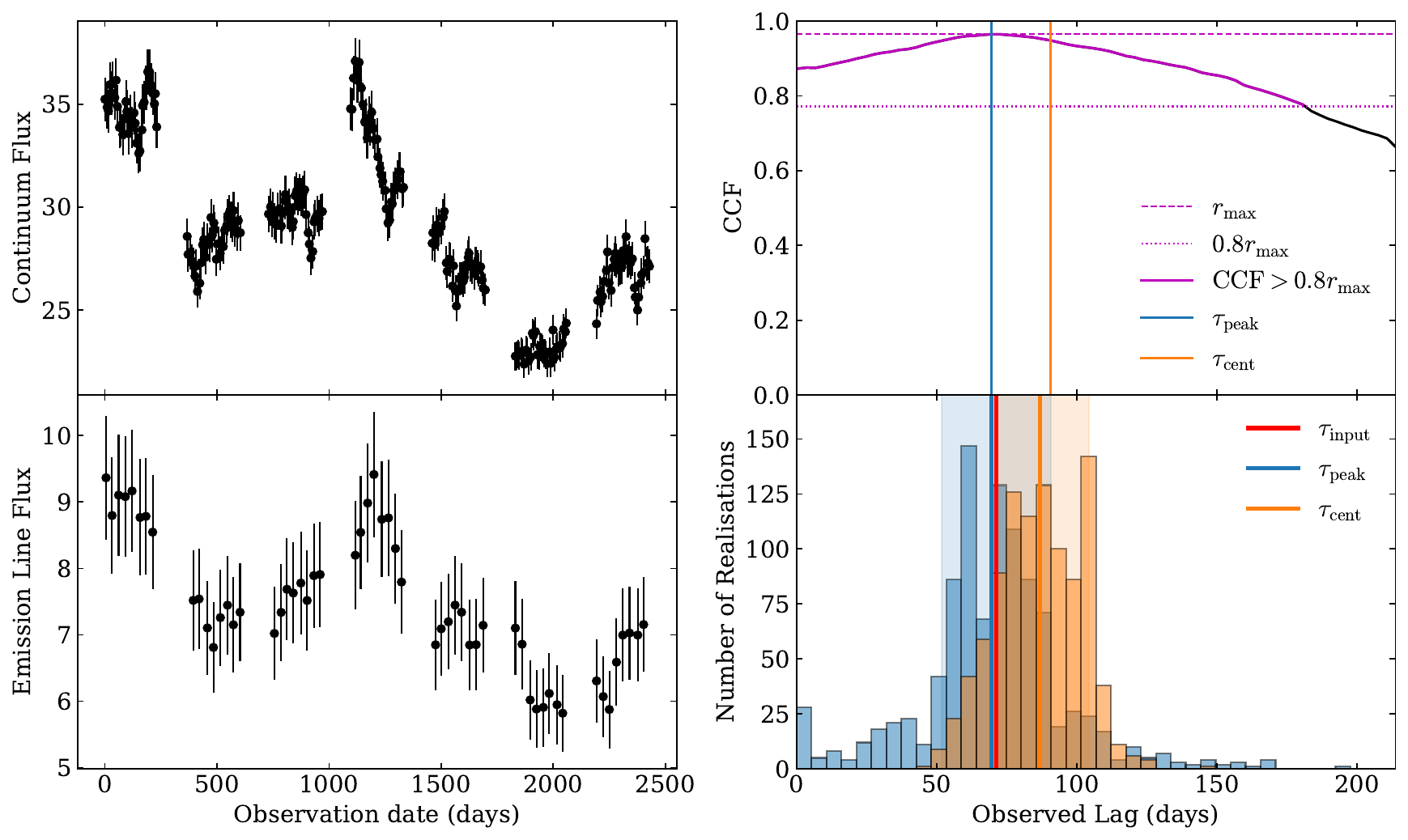}
	\caption{Example of the interpolated cross-correlation function (ICCF), applied to a pair of simulated light curves. The left panels show the simulated continuum and emission-line light curves. The top right panel shows the cross-correlation function computed using ICCF, highlighting the section of the CCF used to compute the centroid, and the peak lag and centroid lag measurement from this CCF. The bottom right panel shows the cross-correlation centroid distribution (CCCD) in orange, and the cross-correlation peak distribution (CCPD) in blue, with the vertical lines and shaded regions indicating the recovered lag and its uncertainty, from each distribution. The red line indicates the input observed-frame lag, $\tau_{\rm input}$, for the source.} 
    \label{fig:ICCF}
\end{figure*}

One of the most widely used lag recovery techniques is the interpolated cross-correlation function \citep[ICCF,][]{Gaskell1987}. In order to focus on the effects of the window function over the critical variables, we chose not to use the more complex \texttt{JAVELIN} \citep{Zu2011} methodology in this work. For comparisons of the ICCF and \texttt{JAVELIN} lag recovery methods, see \citet{Li2019}, \citet{Yu2020}, and \citet{Penton2022}.

We use the ICCF method as described by \citet{Peterson1998}, and summarised here. The recovered lag and its uncertainty are measured using the flux randomisation and random subset sampling (FR/RSS) method. Since the ICCF method does not directly use the flux measurement uncertainties when computing the cross correlation function (CCF), flux randomisation accounts for the uncertainties by shifting the original flux values by Gaussian deviates based on the standard deviation on the fluxes. The random subset sampling accounts for the effect of any one flux measurement on the lag measurements, by randomly selecting light curve points with replacement (resulting in a sample $\sim$37\% smaller than the original set). Both the continuum and emission-line light curve are altered by this FR/RSS process. A CCF is computed with the continuum light curve and the linearly interpolated emission-line light curve, which are cross correlated as a function of time lag. The light curve is linearly interpolated at a specified time lag spacing. Another CCF is computed using the linearly interpolated continuum light curve and the emission-line light curve. The final CCF is the average of these two CCF's. The lag is the time at which the peak of the final CCF occurs. 

A cross-correlation peak distribution (CCPD) is obtained with the lags found from 1000 Monte Carlo realisations of FR/RSS. The recovered lag is taken to be the median of the CCPD, and the uncertainties are the upper and lower limits which contain 68.27\% of the realisations (equivalent to 1$\sigma$ uncertainties of a normal distribution). An example is shown in \autoref{fig:ICCF}. The centroid measurement is also commonly used, as recommended by \citet{Peterson2004}, which is computed as the median of the CCF values $>0.8r_{\rm max}$ counted out from the peak of the CCF. In this work the peak lag was found to be more accurate and consistent than the centroid lag, when dealing with various CCF shapes (which arise due to varied light curve sampling). For example, as shown in \autoref{fig:ICCF}, the centroid measurements were often skewed toward the centre of the lag prior range due to the broad, flat shape of the CCF, and our use of a non-negative lag prior. We investigate the effect of the lag prior in \ref{sec:prior}.

When computing the CCF, we use the PyCCF python code \citep{2018ascl.soft05032S}, setting the interpolation grid spacing to be 3 days. We set the $r_{\textrm{max}}$ threshold to be 0, and disable the automatic use of the $p$-value test to set the $r_{\textrm{max}}$ threshold.

\section{Results} \label{sec:results}

\subsection{Window function parameter space} \label{sec:WFspace}

There are three critical components to the observational window function: observation cadence, the length of the seasonal gap, and the survey duration. In principle the cadence of photometric and spectroscopic data can be varied independently. However initial experimentation indicated that, under the assumption that photometric data is highly likely to be available with greater frequency than the associated spectroscopy, the cadence of said photometry has limited bearing on lag recovery. We therefore proceed with a simplified model for observational cadence, varying photometric and spectroscopic observations in lockstep.

To probe the observational window parameter space, we conducted three sets of simulations: the first keeping the baseline of the survey fixed (reflecting the 6 year OzDES survey) while varying the cadence and season length, the second keeping the cadence constant (weekly photometry and monthly spectroscopy) while varying the baseline and season length, and the third keeping the season length constant (5 months) while varying the cadence and baseline.

\begin{figure*}
	\centering
	\includegraphics[width=0.9\textwidth]{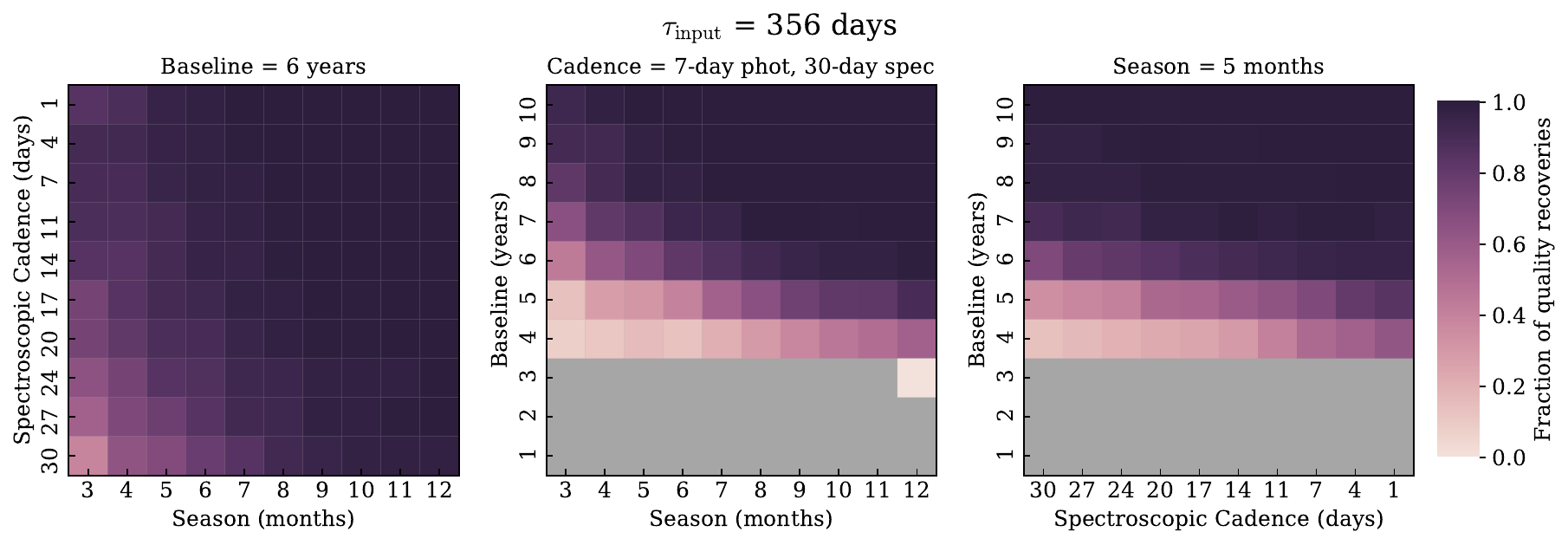}
	\caption{The fraction of quality lag recoveries, across the window function parameter space, for a source with $\tau_{\rm input}$=356\,days. The fixed component of the window function is labelled above each panel. The blank regions are where the baselines of the light curves were shorter than the lag prior range.} 
    \label{fig:WFspace356}
\end{figure*}

\begin{figure}
	\centering
	\includegraphics[scale=0.49]{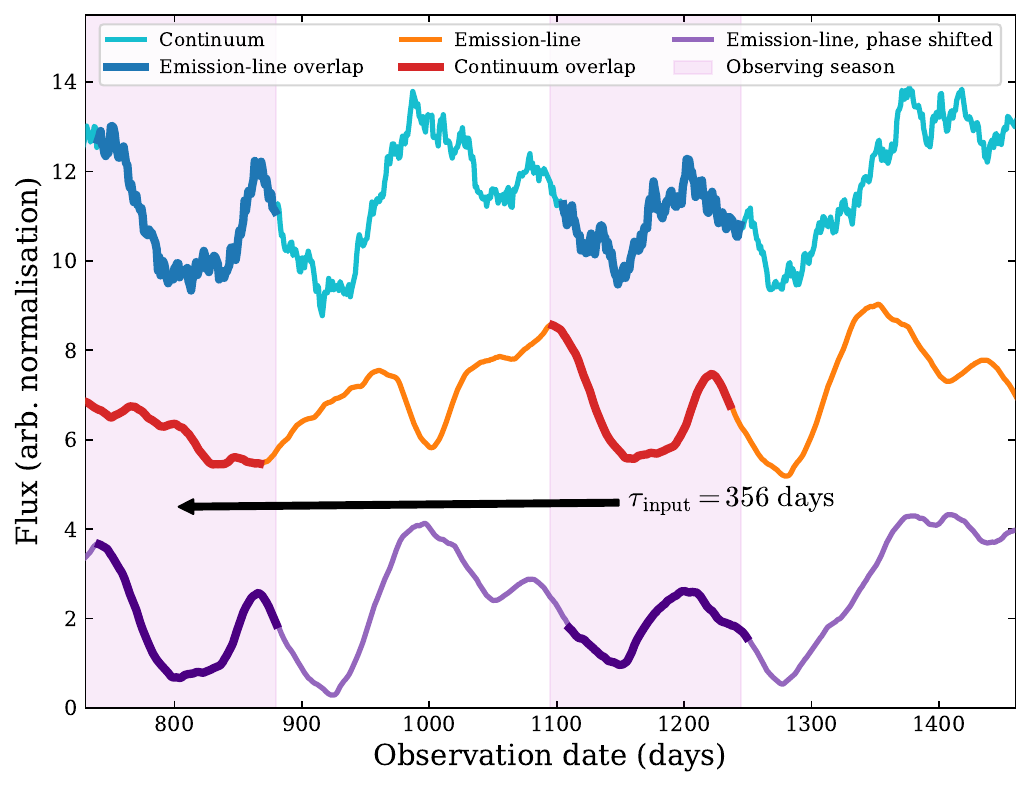}
	\caption{A 2-year section of simulated continuum and emission-line light curves, at daily cadence, with $\tau_{\rm input}$=356\,days. For clarity, the light curve flux errors are not shown here. The phase-corrected emission-line light curve is shown at the bottom. The darkened portions of the light curves indicate where the continuum and emission-line light curves overlap within the observing seasons. As the emission-line light curve lags the continuum by 356 days, the reverberation is seen almost exactly one year later.} 
    \label{fig:LC_356}
\end{figure}

To model various survey cadences the light curves were down-sampled sequentially over 10 levels, ranging from daily photometry and spectroscopy, to weekly photometry and monthly spectroscopy. This last down-sampling level approximately models the DES and OzDES observational cadence. More densely sampled photometry was chosen as it is more readily obtainable and has been shown to improve lag recovery \citep{Shen2015}. The season length was extended by one month at a time, starting from 3 months up to the maximum possible season length of 12 months. The baseline was extended successively by one year, from 1 year to 10 years. Each window function was applied to 100 pairs of light-curve realisations created for a source. 

To best demonstrate the interactions between the lag and various window function timescales, we simulated sources with $\tau_{\rm input}$ $\sim$2 months, $\sim$5 months, $\sim$12 months, and $\sim$18 months.

Our goal is to determine, across the observational parameter space, the fractional recovery of accurate lag measurements with acceptable precision. For each simulated input lag value, we consider the fractional recovery rate for each survey window function realisation. We require a quality lag recovery to be within 10\% (and 100 days) of $\tau_{\rm input}$, and $\geq$ 4 times the statistical uncertainty on the lag measurement provided by ICCF. Following \citet{King2015}, we use a lag prior of 0 to 3$\tau_{\rm expected}$.

\autoref{fig:WFspace356} shows the fraction of quality recoveries for a source with a lag of $\tau_{\rm input}$= 356\,days, through the window function parameter space. For this source with a lag close to one year, there is extensive overlap between the photometry and spectroscopy, albeit from campaign seasons in different years, as shown in \autoref{fig:LC_356}. This lag can be recovered with short seasons and low cadence, with a sufficiently long baseline. Even subtle variability features seen within short seasons will be directly observed in the emission-line light curve. The first results from OzDES \citep{Hoormann2019} were for C\,\textsc{iv} sources with observed lags $\approx1$\,year, as they could be reliably recovered when only the first four years of data were recorded.

\begin{figure*}
    \captionsetup[subfigure]{slc=off}
    \centering
    \begin{subfigure}[b]{0.9\textwidth}
       {\includegraphics[width=1\linewidth]{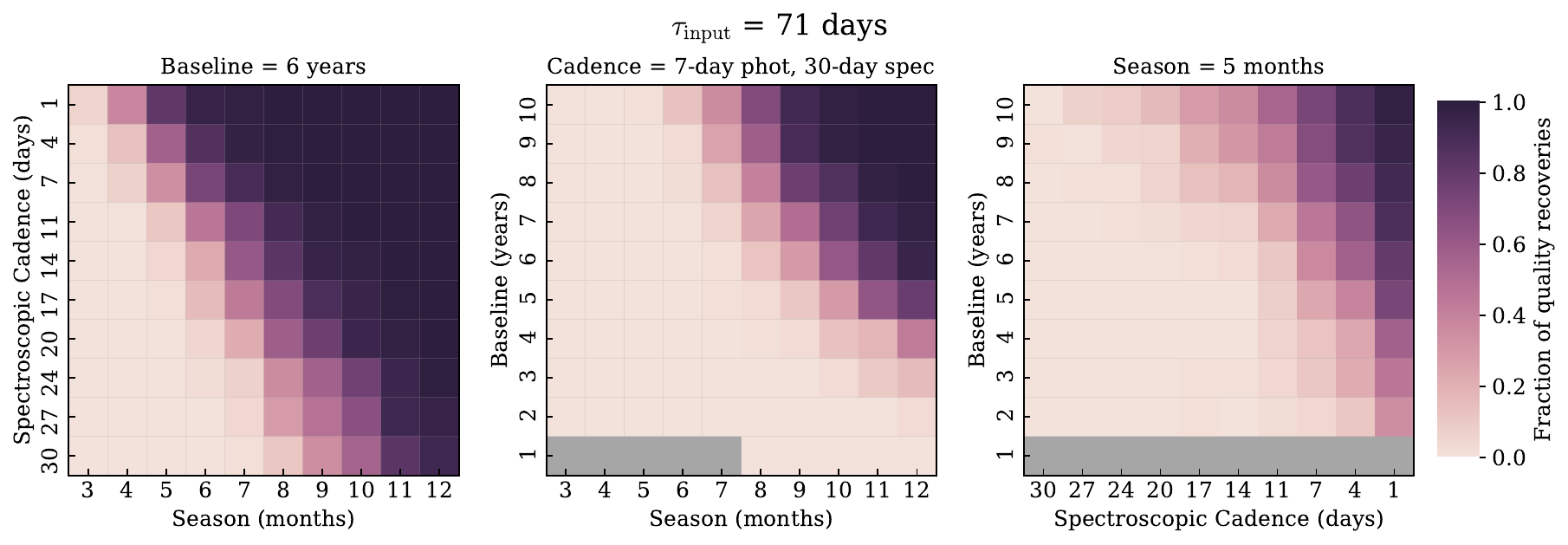}}\\[-10pt]
       \label{fig:Ng1} 
    \end{subfigure}
    
    \begin{subfigure}[b]{0.9\textwidth}
       \includegraphics[width=1\linewidth]{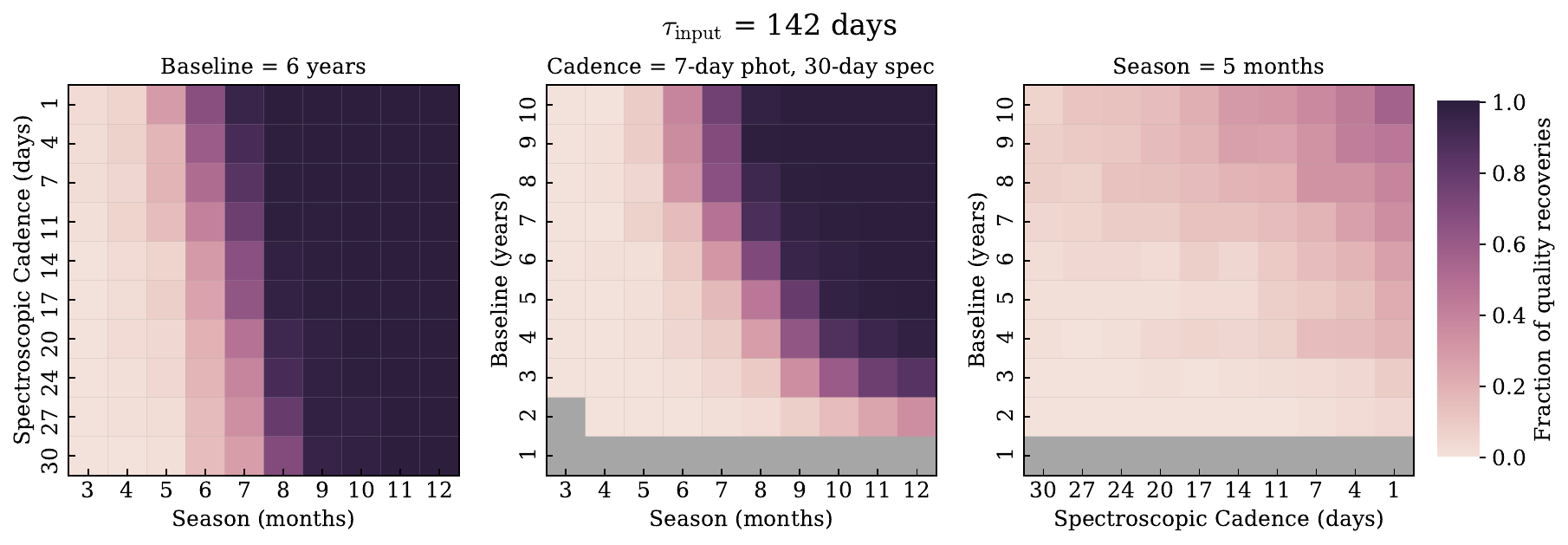}\\[-10pt]
       \label{fig:Ng2}
    \end{subfigure}
    
    \begin{subfigure}[b]{0.9\textwidth}
       \includegraphics[width=1\linewidth]{figs/qualrec_356.pdf}\\[-10pt]
       \label{fig:Ng3} 
    \end{subfigure}
    
    \begin{subfigure}[b]{0.9\textwidth}
       \includegraphics[width=1\linewidth]{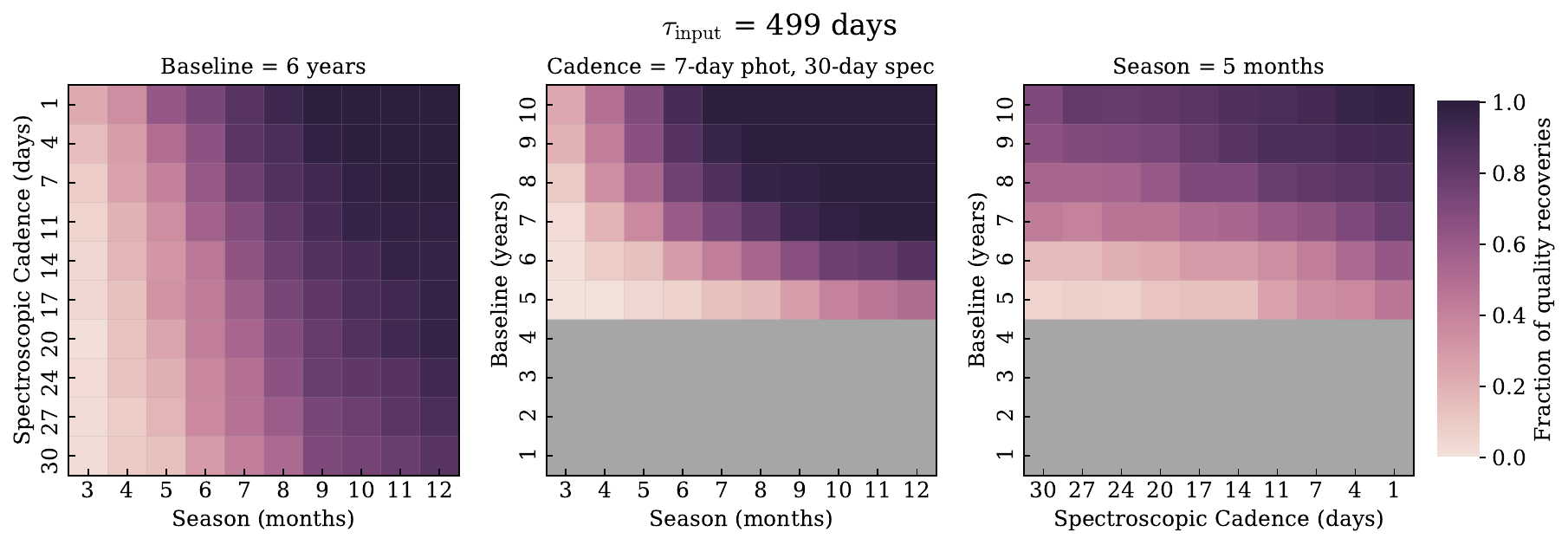}\\[-10pt]
       \label{fig:Ng4}
    \end{subfigure}
    
    \caption{Each row presents the fraction of quality lag recoveries for sources with various input lags, which are labelled above each row, across the window function parameter space. The fixed component of the window function is labelled above each panel. The blank regions are where the baseline of the light curves is shorter than the lag prior range.}
    \label{fig:WFspace}
\end{figure*}

\begin{figure}
	\centering
	\includegraphics[scale=0.49]{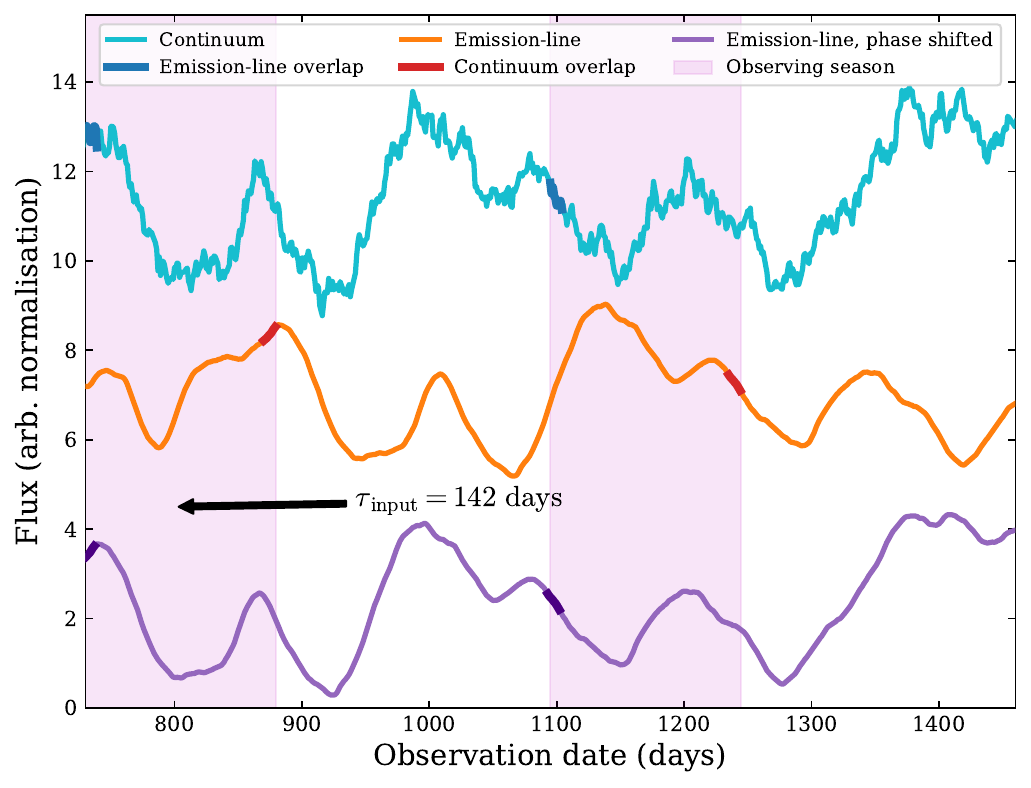}
	\caption{A 2-year section of simulated continuum and emission-line light curves, at daily cadence, with $\tau_{\rm input}$ = 142\,days. For clarity, the light curve flux errors are not shown here. The phase-corrected emission-line light curve is shown at the bottom. The darkened portions of the light curves indicate where the continuum and emission-line light curves overlap within the observing seasons. As the emission-line light curve lags the continuum by 142 days, the direct reverberation response begins when the observing season ends, so it cannot be observed.} 
    \label{fig:LC_142}
\end{figure}

We repeat this for sources with various input lags. \autoref{fig:WFspace} shows the fraction of quality recoveries for a selection of sources with various observed-frame lags, through the window function parameter space. For a source with an input lag of $\tau_{\rm input}=142$\,days, there is significant improvement in the fraction of quality recoveries as the seasonal gap is reduced. As this lag is almost 5 months in the observed frame, it is challenging to recover when the season length is equal to or shorter than the lag. The direct response of the broad-line region to the variability of the ionising continuum occurs mainly outside of the 5-month observing season, as shown in \autoref{fig:LC_142}. Recovery of a reverberation signal from such a source would therefore rely on higher-order correlation timescales, such as the variability relaxation time, as presented by \citep[e.g.,][]{Grier2019,Homayouni2020,Yu2021}.

We see similar trends for the source with $\tau_{\rm input}=500$\,days (\autoref{fig:WFspace}). This is expected as this lag coincides with the second seasonal gap. However, the second order effect due to the damping timescale results in the lag being moderately recoverable even with seasonal gaps present, as information is present in the smoother long-term variations seen on timescales longer than the lag itself. 

For a source with a lag of $\tau_{\rm input}$ = 71\,days (and with a fixed survey baseline of 6 years) there is significant overlap between the photometric and spectroscopic window function within each campaign season. Unsurprisingly, we find that increasing the observational cadence improves the precision of lag recovery, although it has limited impact on accuracy, which requires closing the seasonal gap (maximising the overlap between photometry and spectroscopy with each campaign season) to achieve significant improvement. The light-echo in the emission-line light curve for this source begins around 71 days, therefore, as with the 142-day lag, we require a long enough season length to provide sufficient overlap to allow lag recovery. We see the relatively short lag can still be well recovered with the fixed cadence, if there is a sufficiently long baseline and extended season length. As the baseline or season length increase, the chance increases of seeing favourable structure to variability (\textit{i.e.,} significant large flux excursions from the mean on short timescales).

\subsection{Quality vs. Quantity} \label{sec:qvsq}

We extend these simulations to a sub-sample of OzDES sources. These sources were selected based on their expected lags, to cover the range of expected observed-frame lags for this sample. As most of our sources have expected observed-frame lags up to 1000 days (\autoref{fig:ozdes_AGN}), we set this as the upper limit for these simulations. We apply window functions that vary one component at a time, using a constant number of data points. This allows us to directly compare how each component of the window function affects the recovery of a particular lag. We added approximately 90 photometric epochs and 20 spectroscopic epochs to the base OzDES window function, by independently changing one component of the original window function. In addition to applying the base OzDES window function, we conducted three sets of simulations with the following window function changes: only the season extended (from 5 to 9 months), with the baseline extended (from 6 to 10 years), and with increased cadence (from (7,30) to (4,18) days for the (continuum, emission-line) light curves, respectively).. Lags were recovered for each of 100 pairs of light curve realisations created for each simulated source. 

\begin{figure}
	\centering
	\includegraphics[scale=0.57]{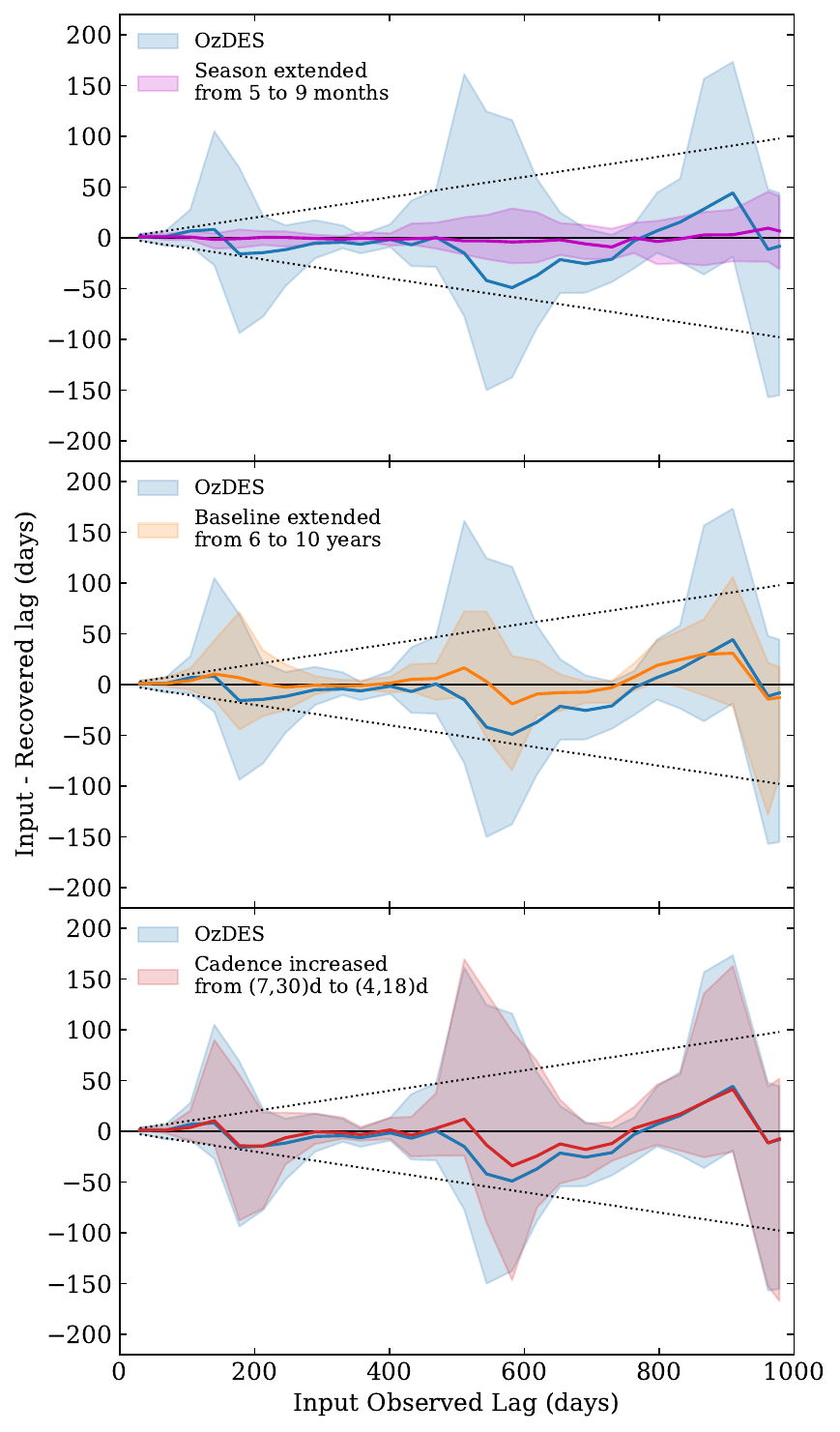}
	\caption{The median (solid lines) and 1$\sigma$ scatter (shaded regions) of recovered lags from 100 realisations, at each input lag. The results of the simulations with the original OzDES window function are shown in blue, along with the results of the following window functions (top to bottom): extended season, extended baseline, and increased cadence. The black dotted lines indicate 10\% of the input lag, which is our criteria for a correct recovery.} 
    \label{fig:qualvsquan}
\end{figure}

The top panel of \autoref{fig:qualvsquan} shows the median and $1\sigma$ scatter of the recovered lags for the simulations in which only the season was extended, compared to the results with original OzDES window function. The same is shown in the middle panel of \autoref{fig:qualvsquan}, for the case with the baseline extended, and in the bottom panel for case with increased cadence. The lags reported here have been recovered from ICCF directly, without any recovery criteria applied.

When sampled by the OzDES window function we see the results expected from the trends observed in \S\ref{sec:WFspace}. There is broad scatter in the recovered lags where the input lag is $n + 1/2$ years long, where $n$ is an integer number of years (\textit{i.e.}, coincident with seasonal gaps). Input lags coinciding with the annual seasonal overlap are recovered accurately, regardless of the window function applied (at the observational data densities considered). The scatter in the lags is significantly reduced across all input lags when the season length is extended. The extension of the baseline reduces the scatter to a greater extent than increasing the cadence. Although the lags are recovered more accurately in these latter cases, a considerable number of lags remain outside the limits for a correct recovery. In these simulations we clearly see the advantage of extending the season length for a large-scale survey hoping to recover lags ranging from tens to hundreds of days. 

\subsubsection{Effect of the lag prior} \label{sec:prior}

\begin{figure*}
	\centering
	\includegraphics[width=0.85\textwidth]{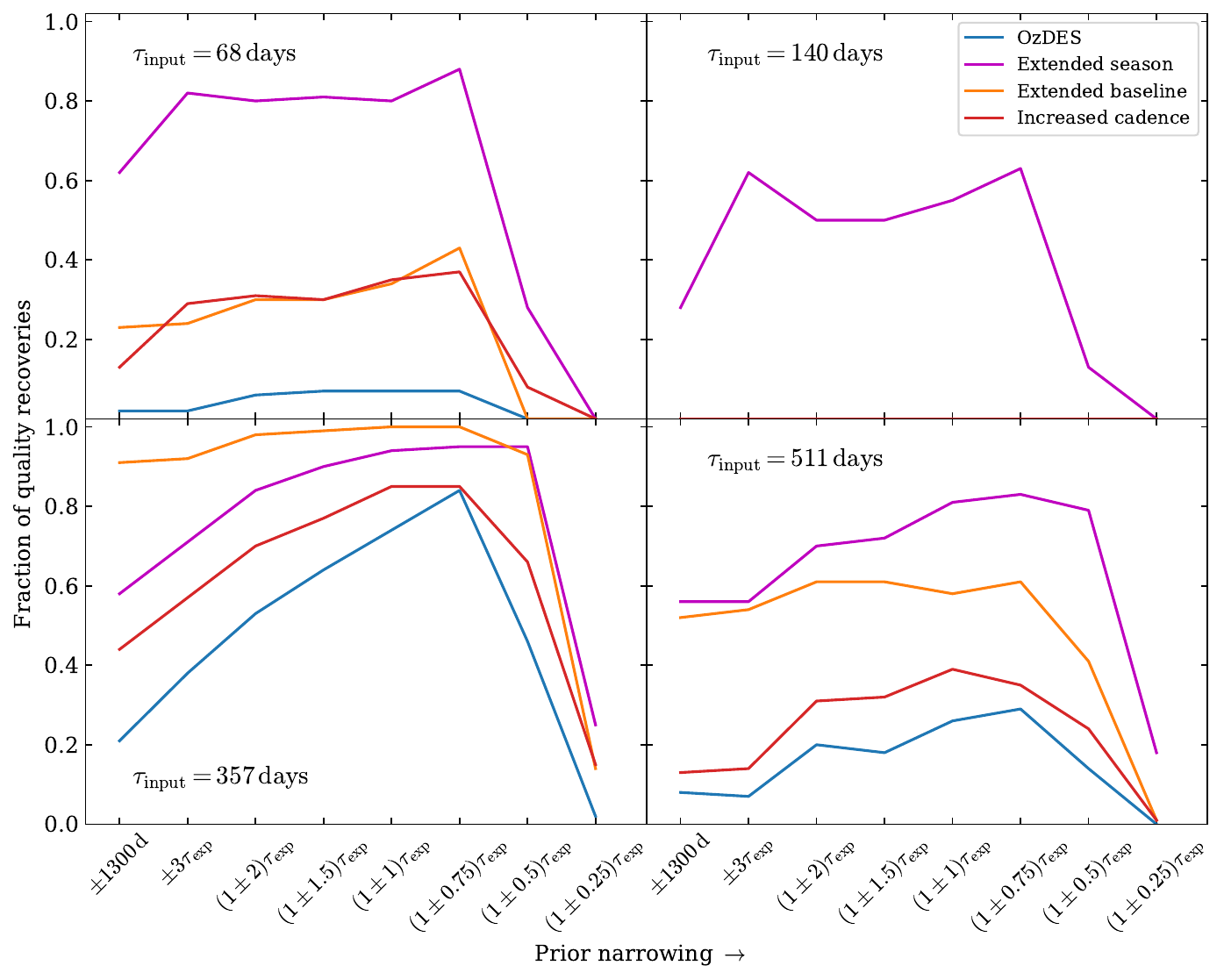}
	\caption{The fraction of quality recoveries for sources with various input lags, which are labelled on each panel, for simulations run with various priors. The results are shown for four different window functions. For $\tau_{\rm input}$=140d, the curves for three of the window functions (OzDES, extended baseline, and increased cadence) are not visible as they have zero fraction of quality recoveries with all lag priors. } 
    \label{fig:priors}
\end{figure*}

There is little consistency within the literature for the priors placed on lag recovery in RM analyses. Whilst there is no objective way to choose a prior, they need to be wide enough to avoid confirmation bias, but also not so wide as to lose the lag in noise or aliasing signals. We investigate the effect of changing the lag prior on lag recovery. Our initial simulations used $0 \rightarrow 3$ times the expected lag $\tau_{\rm exp}$ (which is the same as the $\tau_{\rm input}$ for our simulations), with a maximum allowed prior of 1300 days (about two-thirds of a 6-year baseline). We repeat the lag estimation process using the following lag priors, in order of most broad to most narrow:

\begin{itemize}
    \item $\pm$ 1300 days
    \item $\pm 3 \tau_{\rm exp}$, with a lower bound of -1300 and upper bound of 1300 days.
    \item $(1 \pm 2) \tau_{\rm exp}$, with a lower bound of 0 and upper bound of 1300 days.
    \item $(1 \pm 1.5) \tau_{\rm exp}$, lower bound of 0, upper bound of 1300 days.
    \item $(1 \pm 1) \tau_{\rm exp}$, lower bound of 0, upper bound of 1300 days.
    \item $(1 \pm 0.75) \tau_{\rm exp}$, lower bound of 0, upper bound of 1300 days.
    \item $(1 \pm 0.5) \tau_{\rm exp}$, lower bound of 0, upper bound of 1300 days.
    \item $(1 \pm 0.25) \tau_{\rm exp}$, lower bound of 0, upper bound of 1300 days.
\end{itemize}

\autoref{fig:priors} shows the fraction of quality recoveries with the above priors, for four sources with various input lags. The sources chosen here have lags similar to the sources used for the simulations in \autoref{fig:WFspace}.
Although a prior of $(1 \pm 0.75) \tau_{\rm exp}$ maximises the recovered fraction in most cases, this is very narrow and would result in confirmation bias in RM studies. Overall, when a window function is applied that efficiently recovers a particular lag, it is recovered well and the recovery is mostly unaffected by the prior. This becomes more complex when a lag can be recovered with a certain window function, but less effectively with other window functions. In this case, the recovered fraction can vary greatly depending on the prior.

\subsection{Redshift-Luminosity parameter space} \label{sec:para_space}
 
We now consider the extended parameter space of observations required for cosmological studies. We generate a simulated grid of AGN \citep[similar to][]{Shen2015,King2015,Li2019}, with $16 < m_{r}({\rm AB}) < 23$ over the redshift interval $0<z<5$. The redshift-luminosity range slightly exceeds the range covered by the OzDES and SDSS-RM samples but is inline with what is expected from future generation surveys \cite[e.g.,][]{Kollmeier2017,Brandt2018, Swann2019, MSE2019}. There are also redshift regions where more than one emission line is visible. These regions are particularly valuable as multiple lags can be recovered independently for one source, allowing the ionisation stratification of the BLR to be studied. Future surveys conducted with the Maunakea Spectroscopic Explorer \citep[MSE,][]{MSE2019} will have access to both the optical and near-infrared bands ($J$ and $H$). This will allow observation of the H$\beta$ and Mg\,\textsc{ii} lines at higher redshifts due to the extended wavelength coverage, and increase the range over which two or more emission lines are visible simultaneously for a source. We conduct separate simulations for each emission line, over the redshift range of the grid for which the emission line will be visible with future optical and near-IR surveys. We simulate a pair of light curves for each of 100 AGN per bin, with the expected lag for the relevant emission line.

\begin{figure}
	\centering
	\includegraphics[scale=0.38]{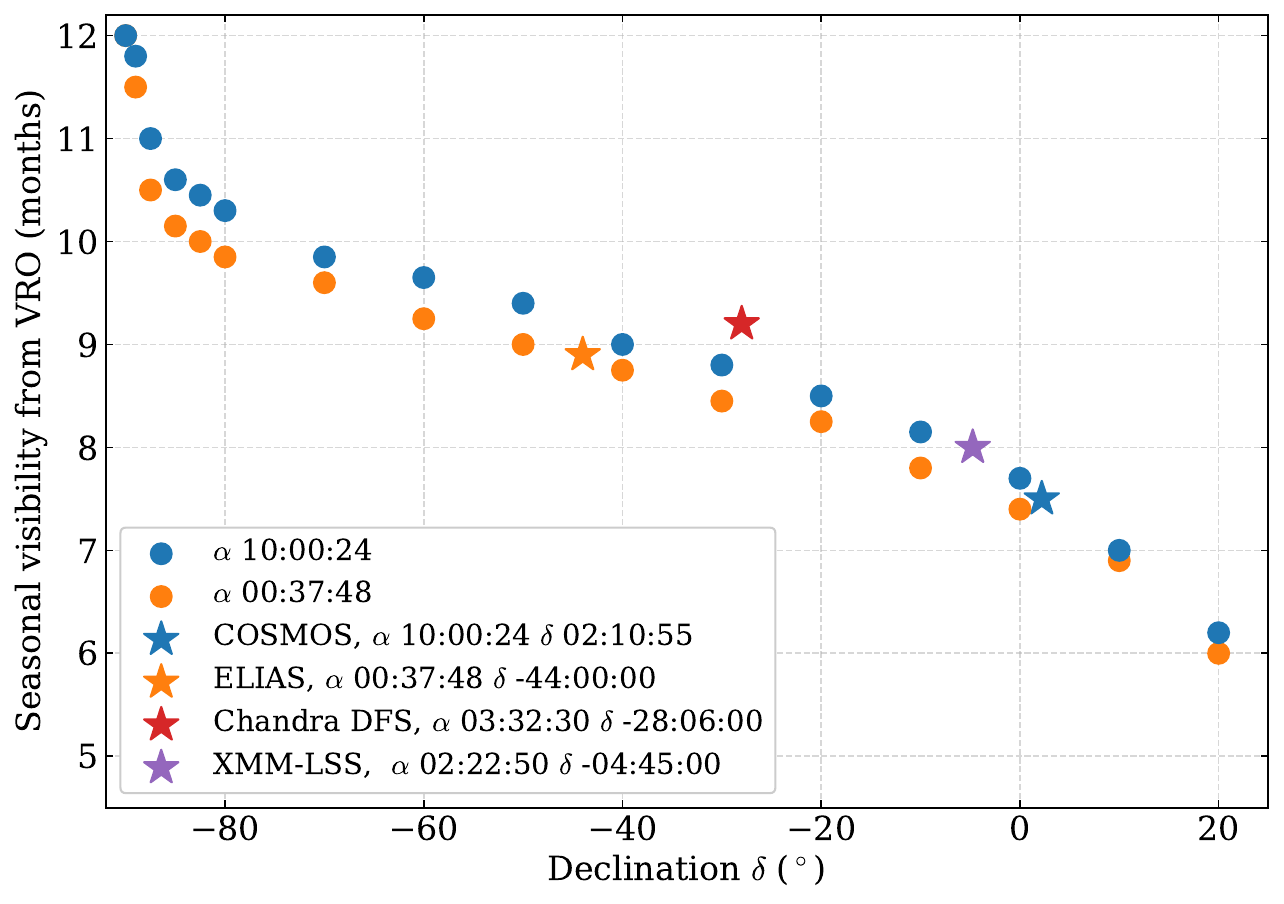}
	\caption{The maximum seasonal visibility at or above an airmass of 2.0 for the four proposed deep-drilling fields for LSST, from the Vera Rubin Observatory at Cerro Pach\'on (-30.2$^{\circ}$). The fields are required to be visible for at least half an hour between astronomical (-18$^{\circ}$) twilights. For reference, the visibility is also shown for hypothetical fields with various declinations, for two right ascensions. At a declination of -90$^{\circ}$, fields would be observed at an airmass of 2.0 year-round from this site. } 
    \label{fig:seas_vis}
\end{figure}

We apply a representative window function of the upcoming LSST and TiDES RM survey to each AGN across the grid. The 5-year baseline planned for 4MOST/TiDES survey \citep{Swann2019} will provide 14$\,\pm\,$4 days spectroscopic sampling, combined with the VRO/LSST RM program \citep{Brandt2018} which proposes to deliver a photometric cadence of $\sim$2 days for each of the four deep-drilling fields. LSST will be conducted using the Rubin Observatory at Cerro Pach\'on (-30.2$^{\circ}$), and TiDES by the VISTA telescope at the Paranal Observatory (-24.6$^{\circ}$). \autoref{fig:seas_vis} shows that the Chandra DFS and ELIAS fields, with the most southerly declination, are visible for the longest maximum season of $\sim$9 months. The equatorial COSMOS and XMM-LSS fields are visible for a maximum of 7.5 and 8 months, respectively. Using the shortest season, we adopt for our baseline survey the window function with a 5-year baseline, 7-month season length, photometric cadence of 2 days and spectroscopic cadence of 14$\,\pm\,$4 days.

\begin{figure*}
	\centering
	\includegraphics[width=\textwidth]{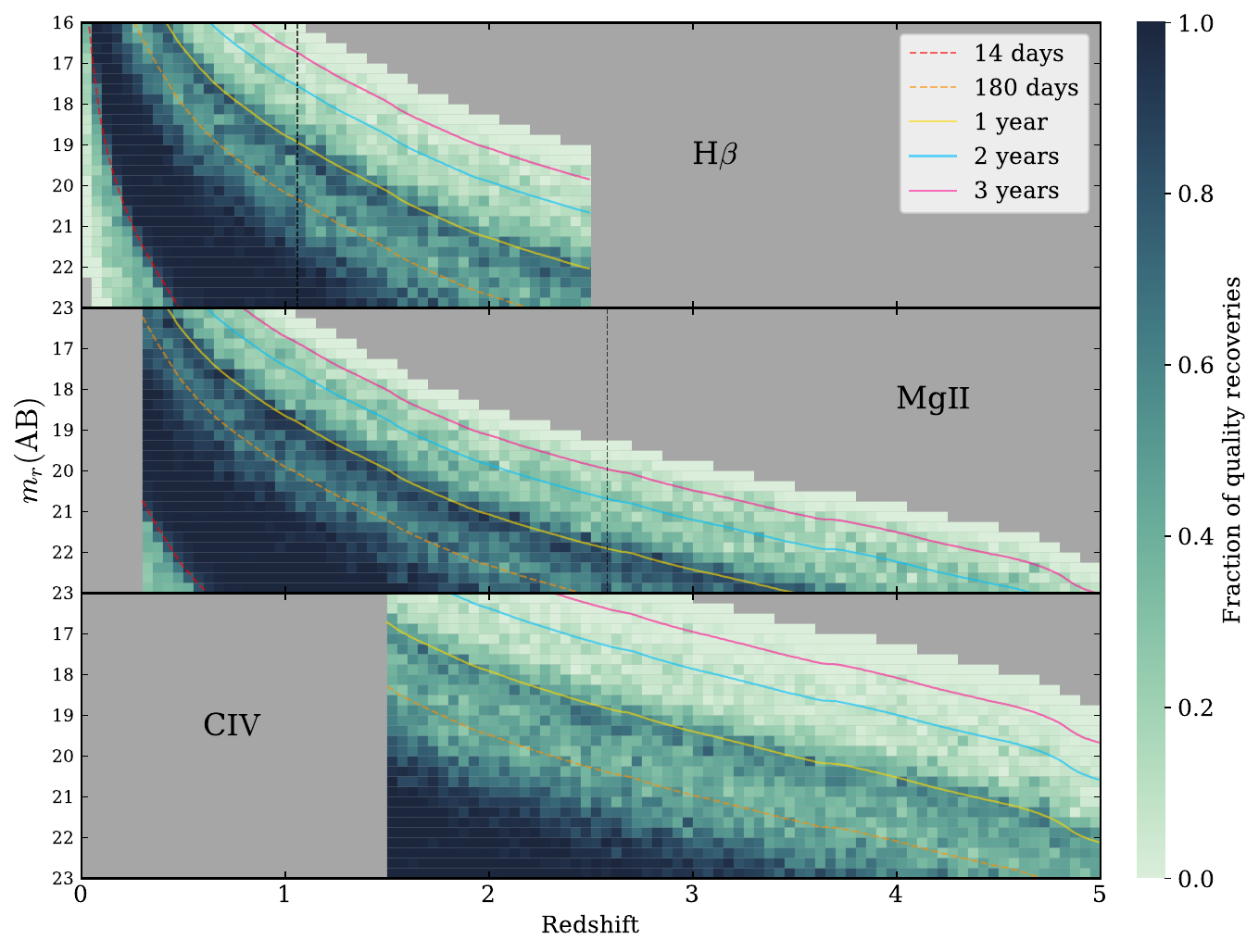}
	\caption{The fraction of quality recoveries across the AGN redshift-luminosity space, with light curves sampled with a representative LSST and TiDES survey window function (5-year baseline, 7-month seasons, 2d photometric and 14d spectroscopic cadence). The time-lag contours indicated by the coloured lines show constant expected observed-frame lag through this parameter space, calculated using published $R-L$ relations for each emission line \citep[][]{Bentz2013,Trakhtenbrot2012,Hoormann2019}. The upper limit on the prior is 1500 days, sources with input lags above this have been masked. For reference, the black vertical lines indicate the redshift at which the respective emission lines are visible at a wavelength of 1 micron (10000\,\AA); representing the transition between visible light and infrared for spectroscopic capabilities. The sensitivity of spectrographs drop off significantly and systematic errors increase near the edges of the wavelength region. Note that this effect is not considered in these simulations.} 
    \label{fig:pop_line}
\end{figure*}

As previously noted, we define a quality lag recovery to be $\geq$ 4 times the statistical uncertainty of the lag measurement provided by ICCF, and within 10\% and 100 days of the input lag. The fraction of quality lag recoveries, for this representative LSST/TiDES window function, \red{is} shown in \autoref{fig:pop_line}, as a function of apparent magnitude and redshift, bounding the probable observable survey parameter space.

We see a banded structure in this {\it visibility map} with high recovery fractions achieved only for sources with observed lags that are out of phase with the seasonal gaps in the window function. Our grid has a resolution of 0.25 in $m_{r}({\rm AB})$ and 0.05 in $z$, which is significantly higher than previous simulations. This was necessary to resolve the high-frequency structure. The highest fraction of quality recoveries are for lags in the region with expected lags between 14 and 180 days, \textit{i.e.,} when the lag occurs within a single campaign season. This lower limit is set by the spectroscopic cadence of 14 day, and the upper bound is the length of the season, which is $\sim$210 days. Beyond this timescale, it becomes challenging to reliably recover lags of longer duration until we approach lags on order of one year, at which point the photometric and spectroscopic observations come back into phase. The pattern is repeated for longer observed-frame lag times at 2 and 3 years, although these bands are fainter due to the limited baseline.

\begin{figure*}
    	\centering
    	\includegraphics[width=\textwidth]{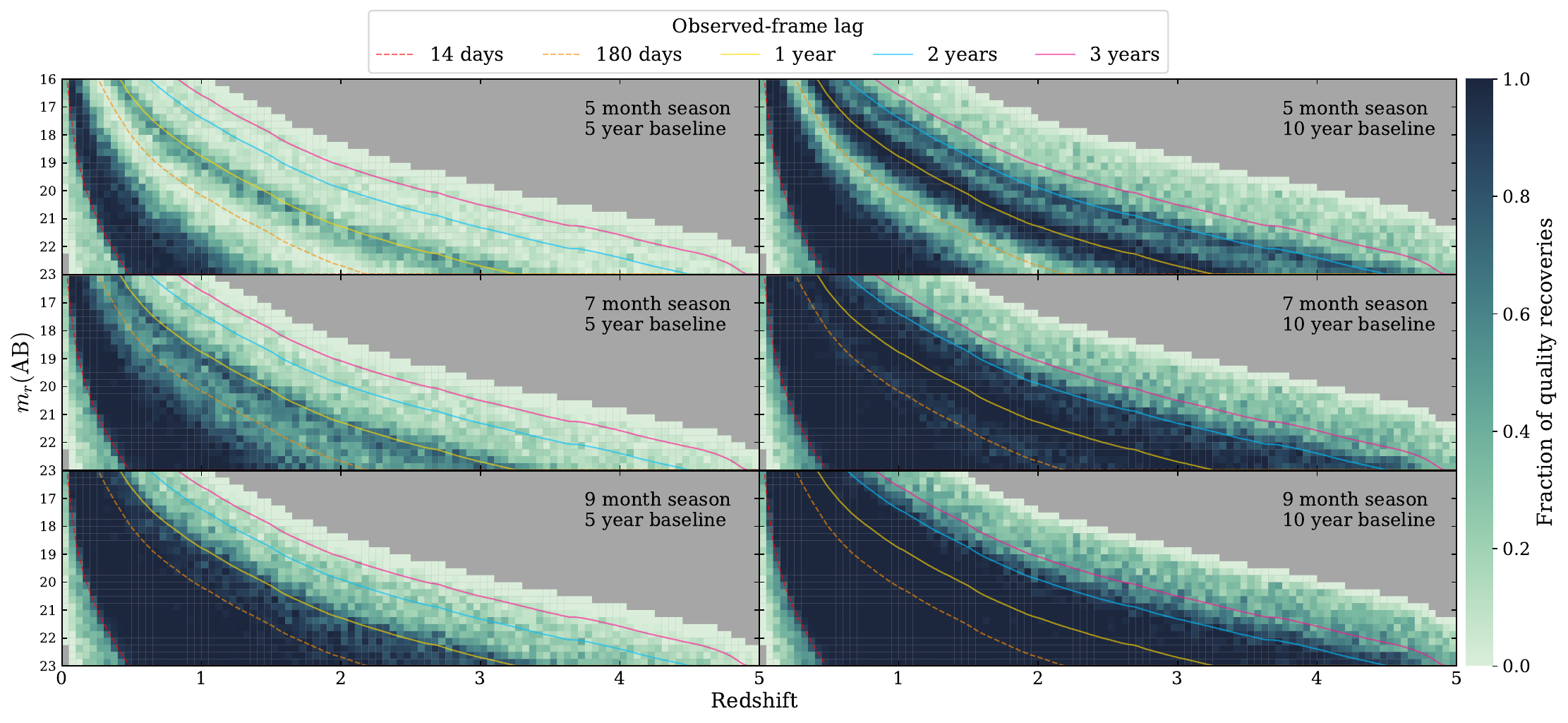}
    	\caption{The fraction of quality recoveries across the AGN redshift-luminosity space for H$\beta$ lags, with light curves sampled by the LSST/TiDES window function labelled on the respective panel. The time-lag contours indicated by the coloured lines show constant expected observed-frame lag through this parameter space, calculated using the \citet{Bentz2013} $R-L$ relation. Sources with input lags above the respective prior limits have been masked.} 
        \label{fig:popHbeta}
    \end{figure*}

\begin{figure*}
    	\centering
    	\includegraphics[width=\textwidth]{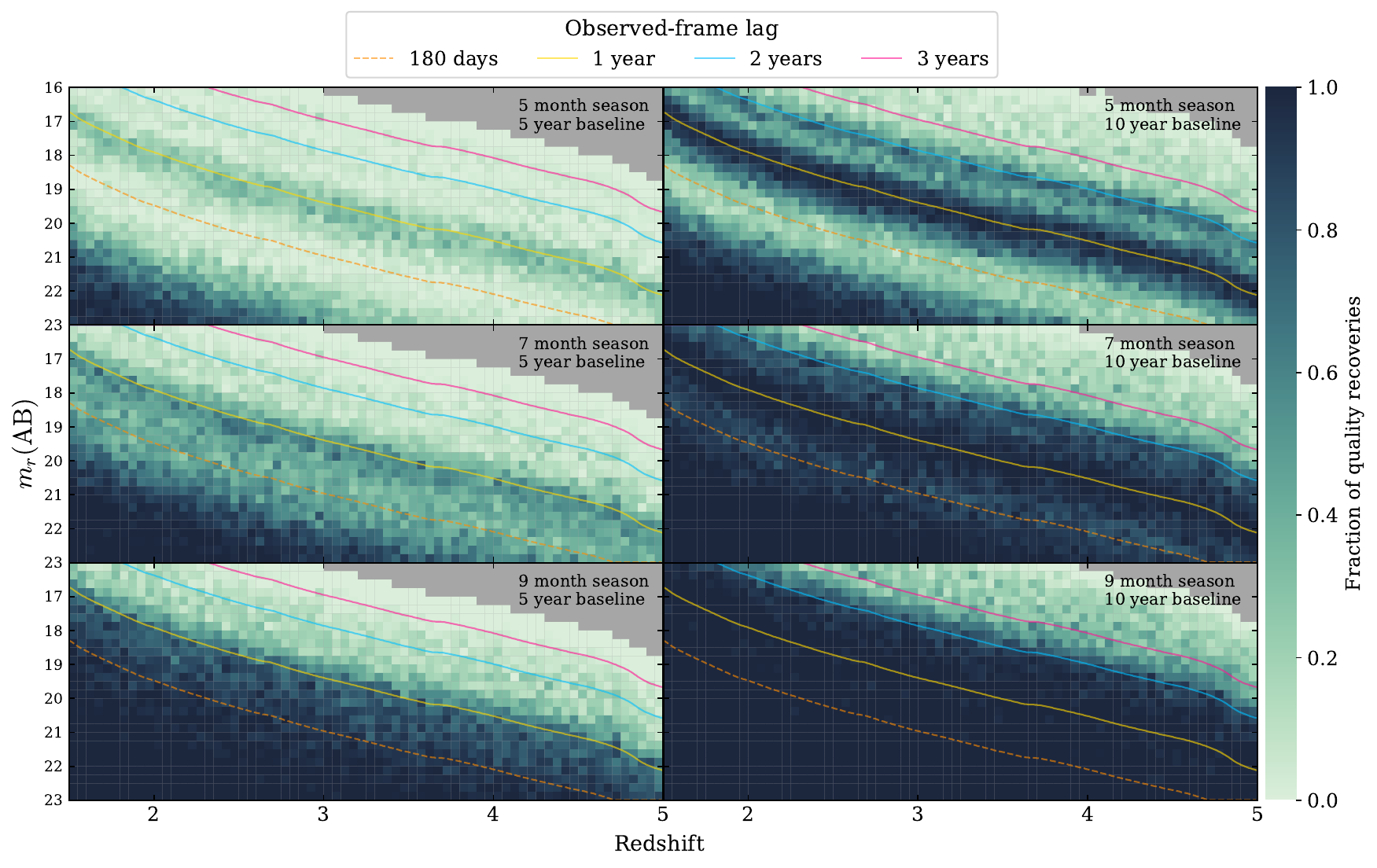}
    	\caption{The same as \autoref{fig:popHbeta} shown for C\,\textsc{iv} lags. The time-lag contours indicated by the coloured lines show constant expected observed-frame lag through this parameter space, calculated using the \citet{Hoormann2019} $R-L$ relation. Sources with input lags above the respective prior limits have been masked.} 
        \label{fig:popCIV}
    \end{figure*}

We have shown in \autoref{fig:WFspace} that extending the season length (closing the seasonal gaps in the window function) generates the greatest improvement in lag recovery rate, followed by extending the survey baseline (\S\ref{sec:qvsq}). We repeated the simulations with three season lengths (5, 7 and 9 months) and two baselines (5 and 10 years). In \autoref{fig:popHbeta} and \autoref{fig:popCIV}, the panels show the results of these simulations with the various window functions, for H$\beta$ and C\,\textsc{iv} lags. We do not perform simulations for Mg\,\textsc{ii} lags separately as there are few measurements constraining this relation \citep{Trakhtenbrot2012}, with a large scatter in recent results \citep{Homayouni2020}. Although there are few sources with lags measured with both H$\beta$ and Mg\,\textsc{ii} \citep[e.g.][]{Clavel1991,Metzroth2006,Homayouni2020}, the lags are found to be similar. Therefore we extend the range of the H$\beta$ simulations to also cover the range of visibility of Mg\,\textsc{ii} with future near-IR facilities.

By extending the baseline to 10 years, the survey becomes much more sensitive to lags around 1 and 2 years. However, the seasonal gaps are still limiting a significant region of this parameter space. We see that the gap in quality recoveries around $\sim$180 days effectively disappears once we extend the season to 9 months. Finally, when both the baseline and season are extended, we see quality lag recoveries for almost all AGN across the parameter space, with the fraction decreasing only for lags longer than 2-3 years. For these longest lags, the lag recovery could benefit from revising/reducing the lag prior (\S\ref{sec:prior}) or extending the baseline further.

\subsubsection{Survey efficacy and the $R-L$ relation}

\begin{table}
    \centering
\caption{The percentage of quality recoveries for a mock target sample from the visibility maps of each window function, done for each emission line (see \autoref{fig:popHbeta} and \autoref{fig:popCIV}).}
	
	\begin{tabular}{lccccc} \hline
                & \multicolumn{2}{c}{H$\beta$}                         & \multicolumn{2}{c}{C\,\textsc{iv}}                            \\ 
WF            & 5 years & 10 years & 5 years & 10 years \\ \hline
5 months          & 18.6         & 63.3               & 18.9                & 53.0 \\ 
7 months            & 38.5         & 82.7                & 61.0              & 95.2 \\ 
9 months            & 54.5         & 87.7                & 92.7             & 99.8 \\ \hline
\end{tabular}
	\label{tab:WFqualrec}
\end{table}

\begin{table*}
\centering
\caption{The best fit slope and intercept parameters of the $R-L$ relation fits in \autoref{fig:RL_CIV}.}
\begin{tabular}{ccccc} \hline
          & \multicolumn{2}{c}{Full sample}       & \multicolumn{2}{c}{`Visible' sample}  \\
WF (season, baseline) & $\alpha$          & $\beta$           & $\alpha$          & $\beta$           \\ \hline
5 months, 5 years   & 0.476 $\pm$ 0.019 & 0.828 $\pm$ 0.032 & 0.403 $\pm$ 0.029 & 0.911 $\pm$ 0.034 \\
5 months, 10 years   & 0.474 $\pm$ 0.006       & 0.838 $\pm$ 0.009       & 0.483 $\pm$ 0.004       & 0.828 $\pm$ 0.007       \\
7 months, 5 years   & 0.467 $\pm$ 0.01        & 0.854 $\pm$ 0.016       & 0.45 $\pm$ 0.006        & 0.875 $\pm$ 0.009       \\
7 months, 10 years   & 0.491 $\pm$ 0.007       & 0.817 $\pm$ 0.012       & 0.484 $\pm$ 0.001       & 0.841 $\pm$ 0.006       \\
9 months, 5 years   & 0.495 $\pm$ 0.01        & 0.812 $\pm$ 0.017       & 0.476 $\pm$ 0.003       & 0.841 $\pm$ 0.006       \\
9 months, 10 years   & 0.487 $\pm$ 0.001       & 0.823 $\pm$ 0.002       & 0.487 $\pm$ 0.001       & 0.823 $\pm$ 0.002      \\ \hline
\end{tabular}
\label{tab:CIV_RL_param}
\end{table*}

We defined a mock target sample using the redshifts and apparent $r$-band magnitudes of the OzDES RM sample (see \autoref{fig:ozdes_AGN}). Each mock target was randomly selected from the respective redshift-luminosity bin on the simulation grid. We calculated the percentage of sources with a quality lag recovery. \autoref{tab:WFqualrec} shows the mean percentage for each window function simulation. 

\begin{figure*}
    \centering
    \includegraphics[width=0.85\textwidth]{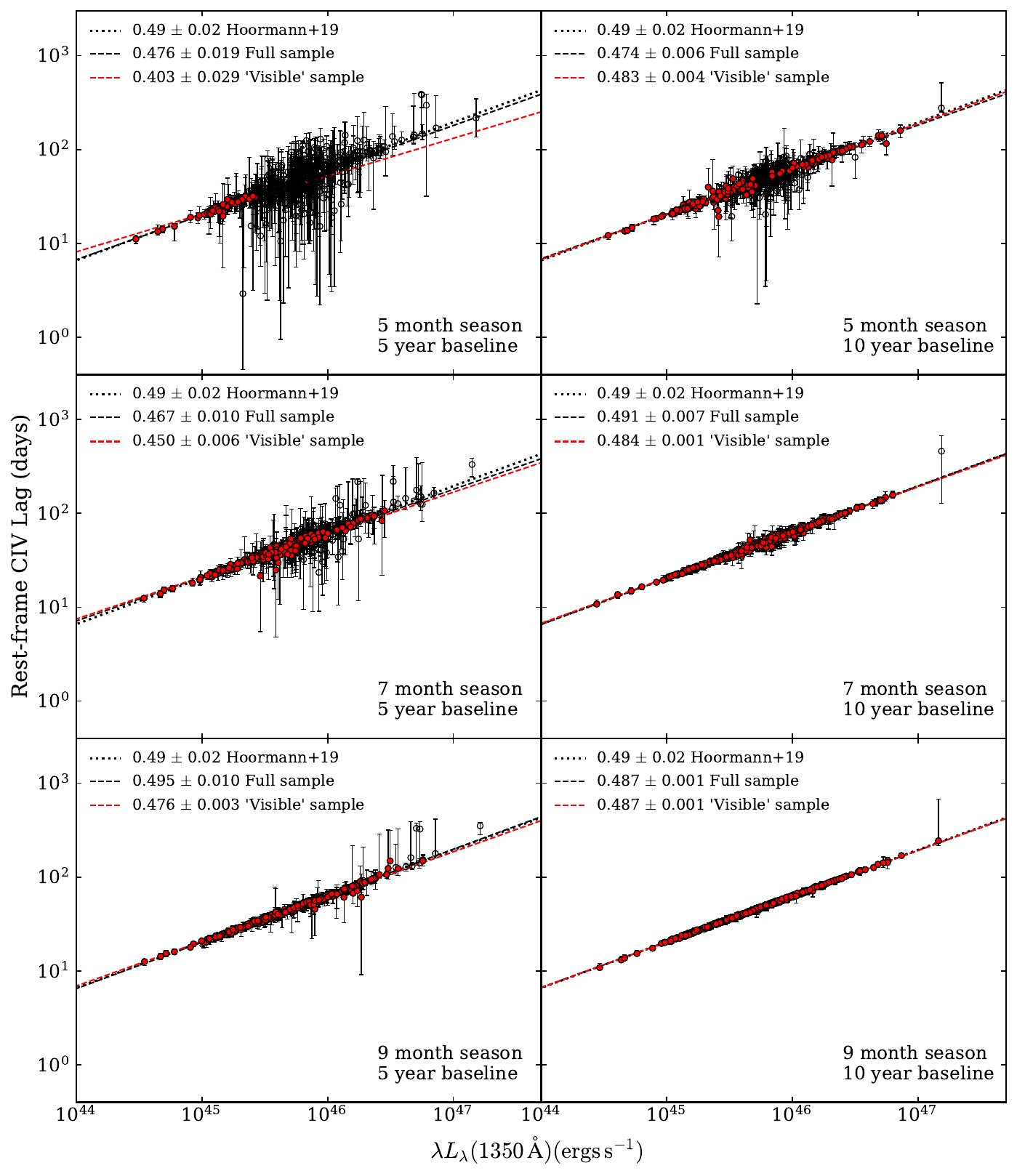}
    \caption{The recovered lags and respective uncertainties for the full mock target sample (black), and `visible' subsample (red), from each window function simulation. The C\,\textsc{iv} $R-L$ relations constrained using the full mock sample and using only the `visible' subsample are shown, along with the input $R-L$ relation from \citet{Hoormann2019}. The slopes for each relation are provided in the legend on each panel. } 
    \label{fig:RL_CIV}
\end{figure*}

The visibility maps created from our simulations show banding structure in AGN luminosity-redshift space for lag recovery efficacy. This identifies regions of this parameter space where reverberation mapping is more likely to be successful, and we use this to inform survey planning. This would require understanding the biases introduced by the specific window function, as certain lags, and therefore certain luminosity-redshift bins, will be over-represented when constraining new $R-L$ relations. We defined a `visible' mock sample comprised of sources in the mock sample selected from a redshift-luminosity bin with an overall fraction of quality recoveries $\geq$\,0.5. The $R-L$ relation was constrained using the full mock sample, and again using only the `visible' mock sample, using the Bivariate Correlated Errors and Intrinsic Scatter (BCES) code \citep{Akritas1996,Nemmen2012}. This result is shown in \autoref{fig:RL_CIV} for each window function of the C\,\textsc{iv} simulations, with the respective $R-L$ relation fit parameters provided in \autoref{tab:CIV_RL_param}. The scatter in the recovered lags from the mock samples is reduced significantly as the season length and baseline are extended. The dynamic range in luminosity of the `visible' mock sample is limited when the season length and baseline are short. As a result of statistical leverage, the slope of the recovered $R-L$ relation is shallower. Although the scatter in the recovered lags is greater for the full mock sample, the slope recovered for the $R-L$ relation is consistent within the fitting error.

% \begin{landscape}
% \end{landscape}

% \begin{landscape}
% \end{landscape}

\section{Discussion} \label{sec:discussion}

\begin{figure*}
    \centering
    \includegraphics[width=\textwidth]{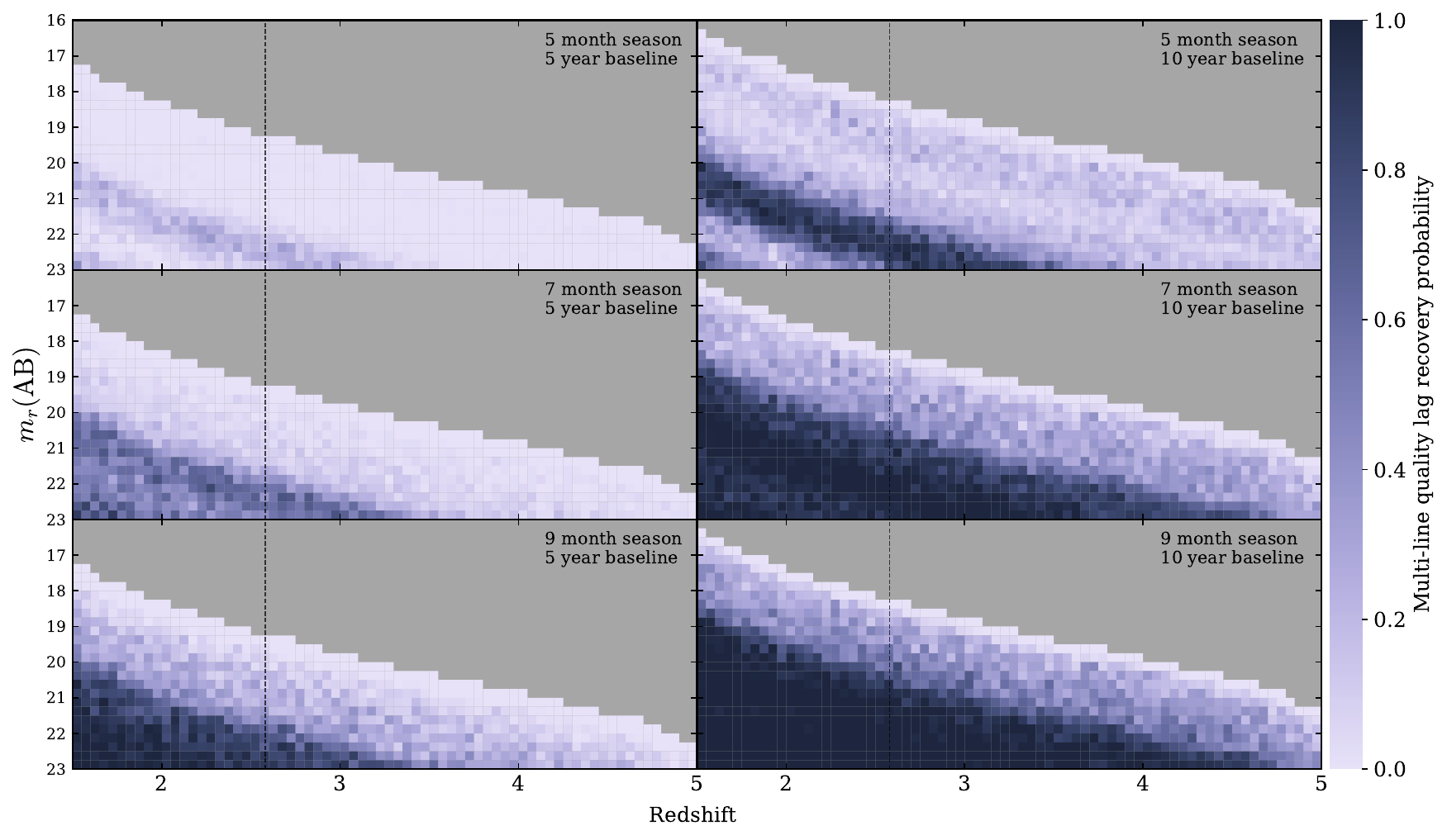}
    \caption{Probability of quality lag recoveries made with both emission lines, for each window function. This was generated by multiplying the visibility maps shown in \autoref{fig:popHbeta} and \autoref{fig:popCIV}. The vertical black lines indicate the maximum redshift range over which either H$\beta$ or Mg\,\textsc{ii} can be seen at visible wavelengths ($0<z\lesssim2.65$). With infrared capabilities, H$\beta$ can be observed up to $z\sim$2.5, and Mg\,\textsc{ii} to $z=5$ (as shown in \autoref{fig:pop_line}); therefore all three lines can be seen from $1.5<z<2.5$. } 
    \label{fig:multi_vismap}
\end{figure*}

It is evident that the window function has a significant effect on the reverberation mapping lag recovery over a range of AGN intrinsic and observed properties. 

\begin{itemize}
    \item \textit{Cadence:} Improvement in the precision of lag recovery gained by increasing the observational cadence is an expected result, as previous studies have also shown \citep{White1994, Shen2015, King2015}. This is particularly pronounced for sources with shorter lag times (for which low observational cadence is not readily compensated through simply extending the survey baseline). However, for sources with longer observational lags, the baseline is critical even if only modestly sampled. This naturally suggests a multi-tiered observational strategy when resources are restricted (typically the case for spectroscopy), similar to that used by SDSS-RM \citep{Shen2015}.

    \item \textit{Baseline:} The effect of the baseline on the ability to recover long lags had been predicted \citep{Shen2015, King2015}, although current surveys are only now reaching sufficient baselines to confirm such assumptions \citep{Shen2019,Hoormann2019}. 
    
    \item \textit{Season length:} While the limits imposed by seasonal gaps had been recognised previously \citep{Zu2011,King2015}, their impact had not been fully appreciated until recent industrial-scale surveys delivered their early results, which required implementing alias removal techniques \citep{Grier2019,Homayouni2020}.
\end{itemize}

This has motivated our work to investigate the interplay of each of these observational window function components in more detail. A large portion of sources observed by current industrial-scale RM surveys have expected observed-frame lags in the sub-year range, as seen in \autoref{fig:ozdes_AGN}, leaving them vulnerable to the impact of seasonal gaps. This is a cause of concern for current and future surveys, which have or are expected to have significant seasonal gaps, as it limits the number of lags that can be reliably recovered.

\subsection{LSST/TiDES}

From the population simulations conducted with the LSST/TiDES window function, we were able to predict the efficacy of this RM survey. The TiDES sample comprises 700 AGN at $r<21$ and $z<2.5$ \citep{Swann2019}. Focusing on this region of the parameter space, we see that the season length imposes critical limitations on the areas where the survey would be able to successfully recover quality lags. We demonstrated the vast improvement apparent with altered window functions where the baseline or seasons were extended. As LSST is planned to operate for 10 years, it is feasible to extend the TiDES program to cover the same baseline. However, the field visibility cannot not be changed, therefore the efficacy in certain redshift-luminosity ranges will be restricted for the equatorial fields. Significant seasonal gaps may not cause as much of an issue for sources with long relaxation times. However, this would require long survey baselines, of at least 6 years; beyond what is currently planned.

As shown in \autoref{fig:seas_vis}, the south celestial pole, and hence circumpolar fields, are visible year-round from the Rubin Observatory, but at a zenith angle of 60$^{\circ}$ (airmass of 2.0). From Paranal Observatory, the south celestial pole lies at a zenith angle of 65$^{\circ}$ (airmass of 2.4). Unless we build facilities further north or south, or at the poles, the high airmass makes it difficult to achieve the signal-to-noise required. In addition, the atmospheric dispersion correction (ADC) is difficult at these high air masses, and many spectrographs do not have an ADC that can reach 60$^{\circ}$. For example, 4MOST corrects only up to 55$^{\circ}$ zenith angle \citep{deJong2019}. The absence of legacy data for circumpolar fields presents a further challenge. Our simulations show that 9-month seasons bridge the seasonal gap sufficiently (e.g., \autoref{tab:WFqualrec}: 99.8\% quality recoveries for CIV mock sample with 10 year baseline and 9 month season), meaning no source is critically dependent on closing the remaining gap of three months.

\begin{itemize}

    \item \textit{Recovering the $R-L$ relation:} Optimising surveys ultimately depends on the science goals. For RM studies this would be not only to recover as many lags as possible for a sample, but also to ensure measurements are made across a range of luminosity, to anchor both ends of the $R-L$ relationship, for each line. The lower luminosity ends of the relations for Mg\,\textsc{ii} and C\,\textsc{iv} still lack measurements, as the expected observed-frame lags are short. The window functions of recent surveys have not been optimised to recover these short lags. As shown in \autoref{fig:RL_CIV}, the baseline length is important for this, particularly when seasonal gaps are present.

    \item \textit{Multi-line measurements:} Other science goals include recovering lags for multiple emission lines, to study the ionisation stratification of the BLR \citep[][]{Dietrich1993,Peterson2004}. Considering the banding structure we see in the visibility maps (Figures \ref{fig:pop_line}-\ref{fig:popCIV}) when there are significant seasonal gaps, we need to see if there are regions of the parameter space that have high likelihood of lag recovery while two (or more) lines are visible. As H$\beta$ and Mg\,\textsc{ii} have similar expected lags, the bands of high efficacy overlap for redshift regions where both lines can be seen. However for Mg\,\textsc{ii} and C\,\textsc{iv}, although both lines can be seen over a wider range in redshift (\autoref{fig:pop_line}), there is less overlap between the bands of high efficacy on the visibility maps. The visibility map showing the probability of quality lag recoveries with both lines is shown in \autoref{fig:multi_vismap}, for each window function. For Mg\,\textsc{ii} and C\,\textsc{iv}, the number of redshift-luminosity bins with a high probability of providing multi-line measurements greatly increases with near-IR coverage.
\end{itemize}

\subsection{Comparing to data}

\begin{figure*}
    \centering
    \includegraphics[width=\textwidth]{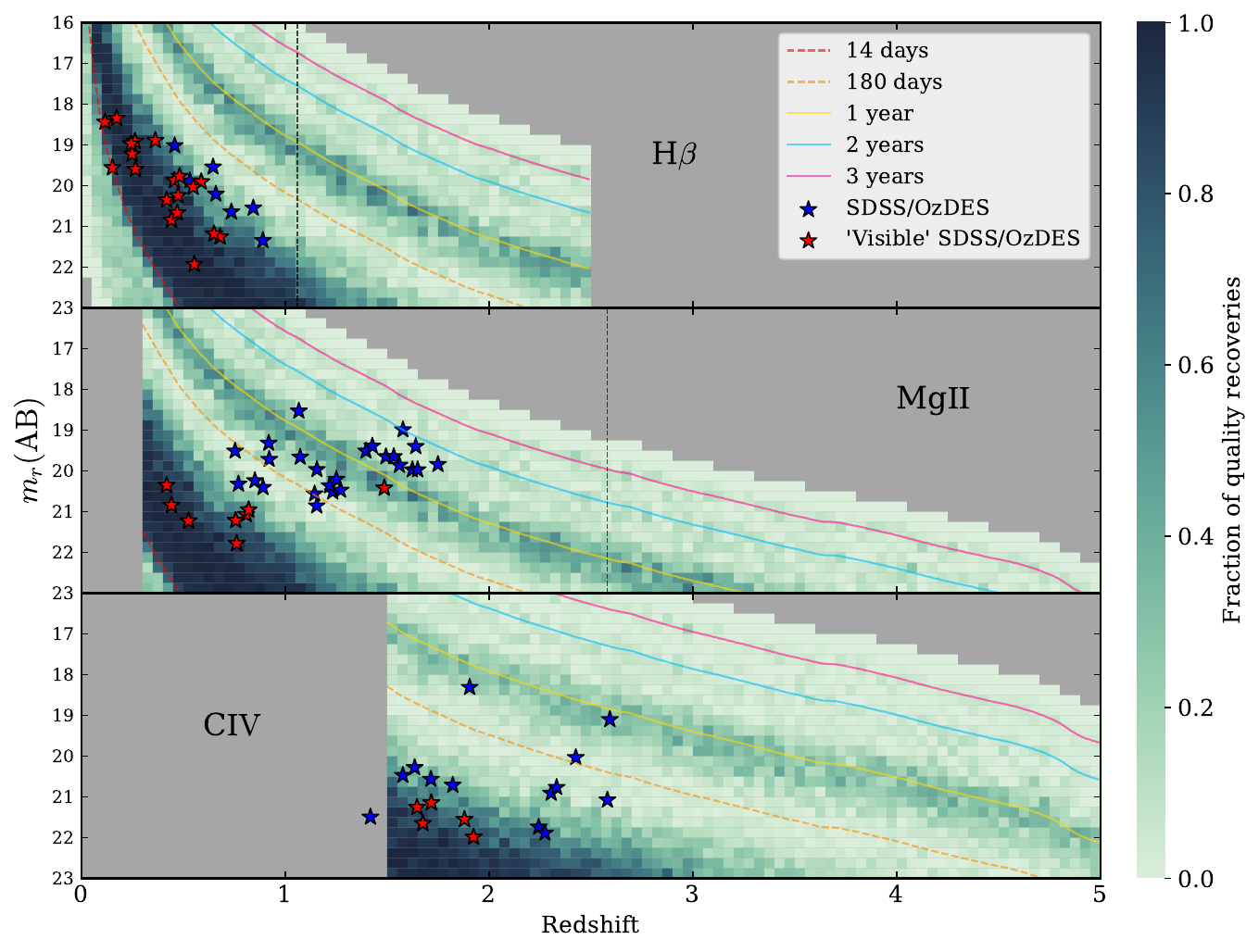}
    \caption{Sources with published high quality lag measurements from the OzDES and SDSS-RM surveys \citep[][]{Grier2017,Hoormann2019,Grier2019,Homayouni2020,Yu2021} overlaid on the respective visibility map for each emission line. The `visible' subsample, lying on bins with a fraction of quality recoveries $\geq$0.5, is in red.} 
    \label{fig:plat_vis}
\end{figure*}

\begin{figure}
    \centering
    \begin{subfigure}[b]{\columnwidth}
        \centering
        \includegraphics[width=0.94\textwidth]{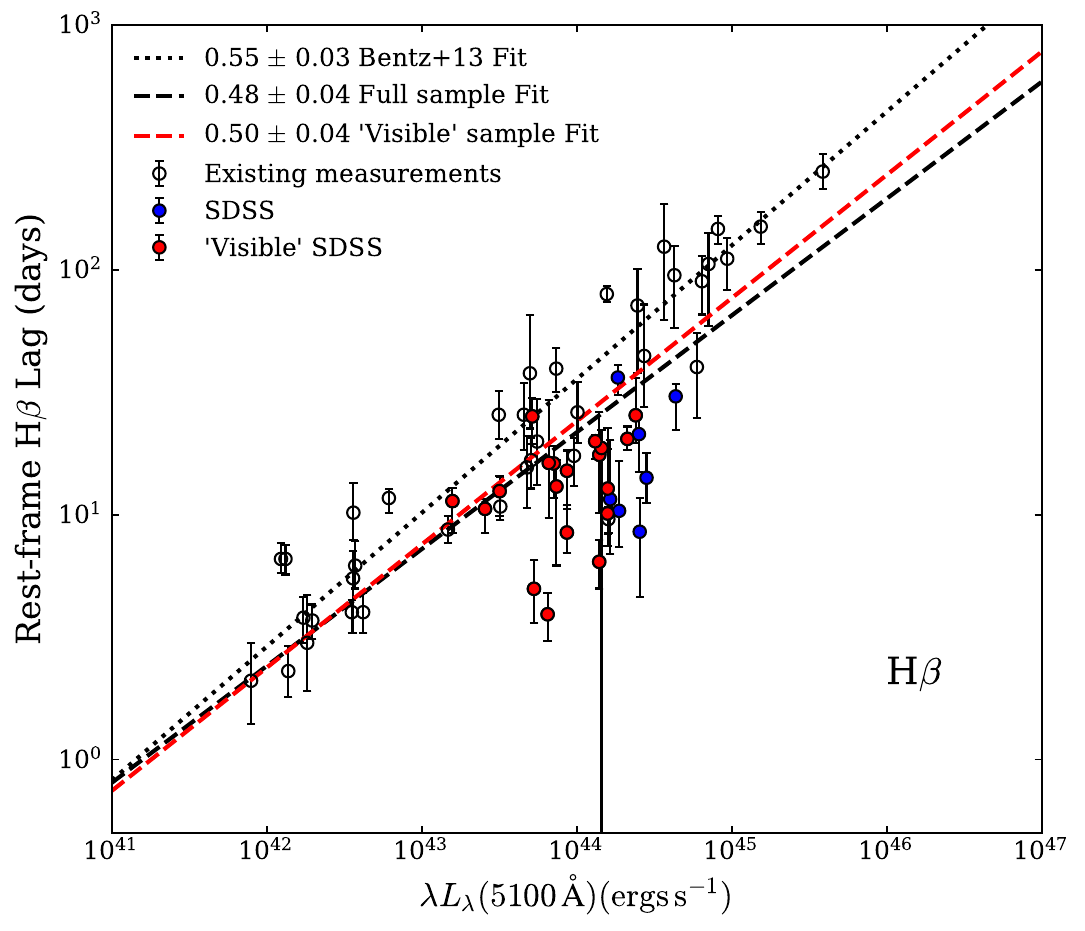}
    \end{subfigure}
    % \hfill
    \begin{subfigure}[b]{\columnwidth}  
        \centering 
        \includegraphics[width=0.94\textwidth]{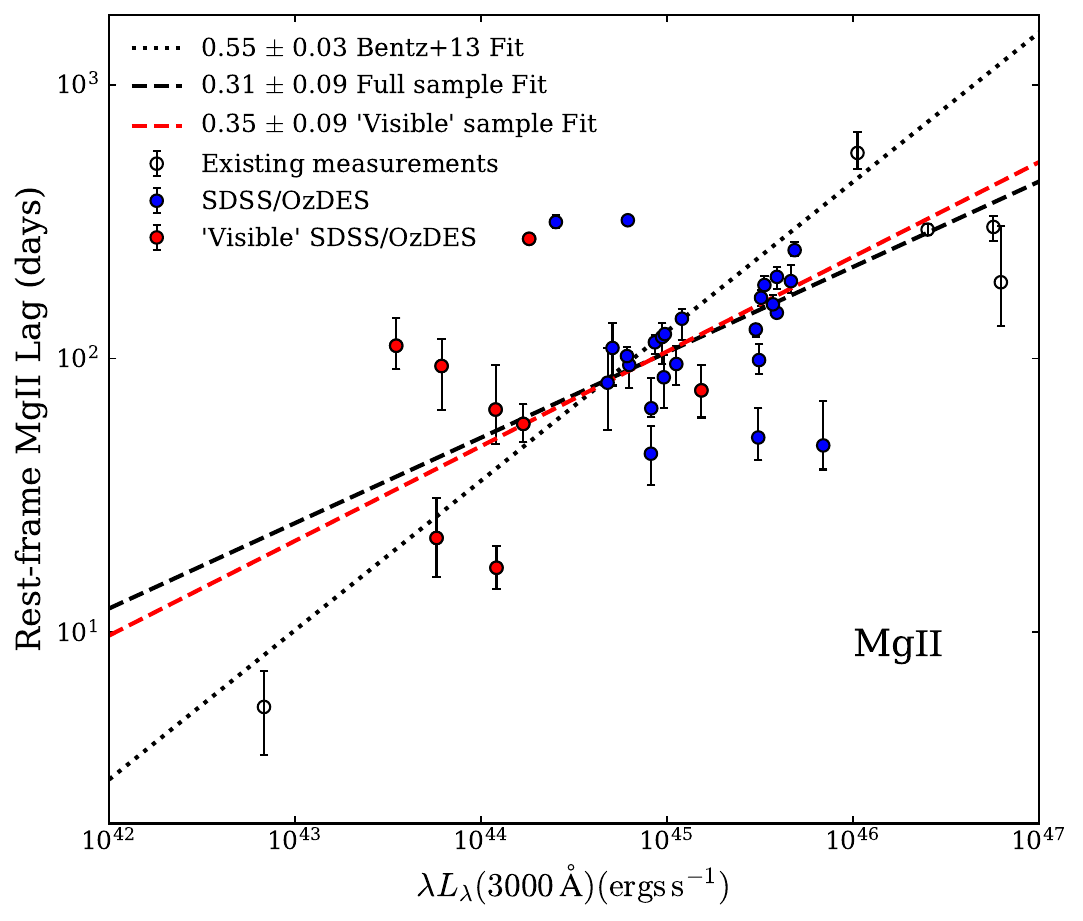}
    \end{subfigure}
    % \vskip\baselineskip
    \begin{subfigure}[b]{\columnwidth}   
        \centering 
        \includegraphics[width=0.94\textwidth]{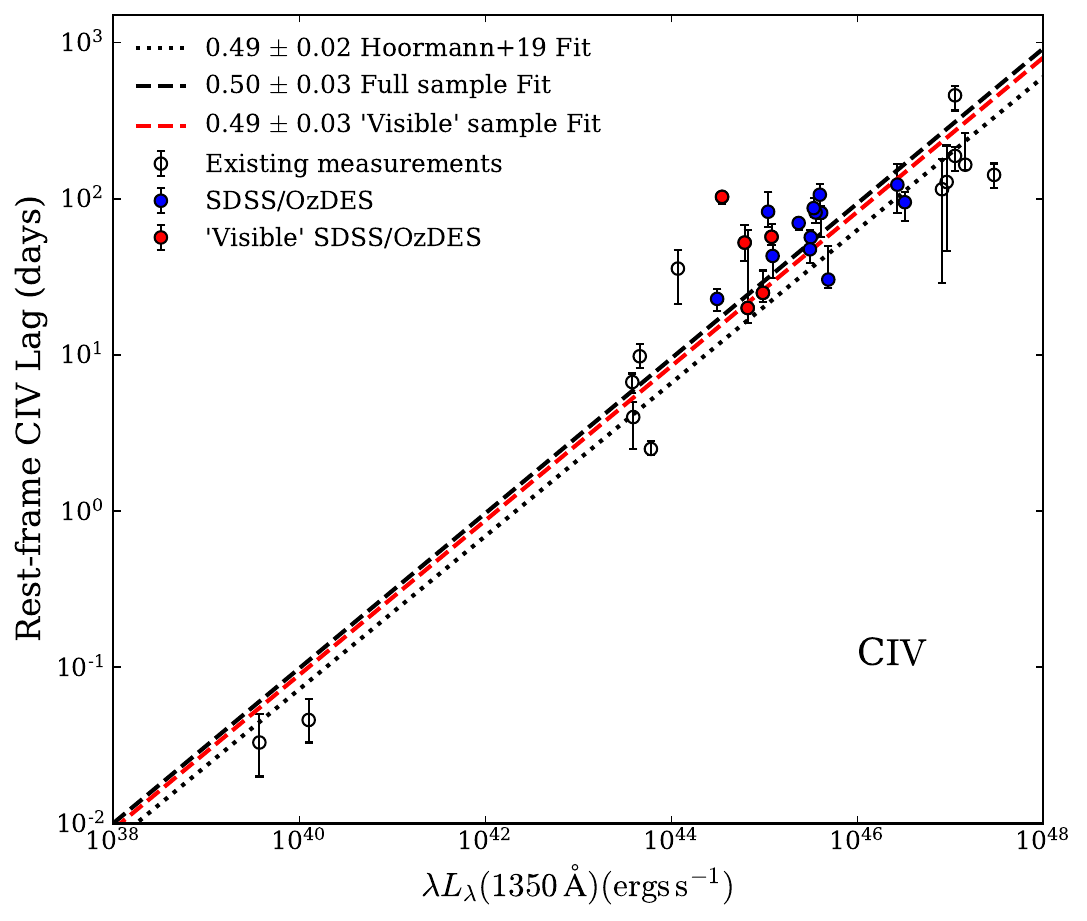}
    \end{subfigure}
    % \hfill
    \caption[]{The $R-L$ relations constrained using the published high quality lag measurements for each emission line (same as in \autoref{fig:plat_vis}), and using only the `visible' subsample. The existing measurements are also used when constraining the relations. These include measurements from \citet{Bentz2013} (and references therein) for H$\beta$; \citet{Metzroth2006,Lira2018,Czerny2019,Zajacek2020,Zajacek2021} for Mg\,\textsc{ii}; and \citet{Peterson2005} (and references therein); \citet{Kaspi2007,Trevese2014,Lira2018} for C\,\textsc{iv}. } 
    \label{fig:plat_RL}
\end{figure}

We compare the results of these simulations to recent lag recoveries from the pioneering  `industrial-scale' OzDES and SDSS-RM surveys \citep[][]{Grier2017,Hoormann2019,Grier2019,Homayouni2020,Yu2021}. In \autoref{fig:plat_vis}, the sources with published high-quality lags from OzDES and SDSS-RM are overlaid on the visibility maps for the respective emission lines. We define a `visible' sample, as previously done for the mock sample, comprising sources which lay on redshift-luminosity bins with a fraction of quality recoveries $\geq$\,0.5. The $R-L$ relations recovered using the full and `visible' published samples are shown in \autoref{fig:plat_RL}. A significantly shallower slope for the Mg\,\textsc{ii} $R-L$ relation was found by \citet{Homayouni2020}, which they suggest could be attributed to the larger intrinsic scatter in the recovered lags. The `visible' samples do not have less scatter, although the number of data points is limited. We note that all the published lags were recovered using \texttt{JAVELIN} \citep[][]{Zu2011}, which could cause some discrepancy with our predictions, which were performed using ICCF. There are many Mg\,\textsc{ii} and C\,\textsc{iv} lags that lie in the seasonal gaps. These sources have been observed to vary smoothly on multi-year timescales, allowing the lag to be recovered with little to no overlap between the continuum and emission-line light curves. Our use of the DRW light curve model may not appropriately represent this variability over longer timescales. Although the seasonal gap lag recoveries are judged to be reliable, we should expect to see a similar number of recoveries that do lie on regions of high `visibility', which is not the case for the current sample of Mg\,\textsc{ii} and C\,\textsc{iv} lags. It is possible that this issue will be alleviated with future results from the longer baseline data from these surveys, which should also help to confirm seasonal gap lags. If the issue remains, it confirms the need for future surveys to ensure the seasonal gaps are reduced as much as possible, to see if this is the cause for the lack of high quality lag measurements from multi-object RM surveys in the regions we predict to be `visible'. 

We acknowledge our simulations are idealised, and the following effects should be appreciated. The broad-line region emission response has been observed to demonstrate non-reverberating behaviour, such as the BLR `holiday' \citep[][]{Goad2016,Dehghanian2019}. The light curve model we used does not include these deviations, as they are still poorly understood. Recently \citet{Stone2022} have found that the DRW damping timescale, $\tau_D$, constrained with 20-year light curves, continues to increase as the baseline of the light curves increases. They also observe smooth, long-term variations in some sources, leading them to suggest that the current DRW model may be insufficient for these sources. The significant seasonal gaps in the window function of the data may also preclude proper convergence of the DRW model parameters. A large sample of emission-line light curves is required to better model reverberation in AGN light curves in future work.

\section{Summary} \label{sec:summary}

We present results of comprehensive simulations which demonstrate the impact of each component of the global observational window function on reverberation mapping lag recovery. From individual sources to a population of AGN, we illustrate the detrimental impact of seasonal gaps. Reverberation mapping surveys should prioritise observing fields with longer seasonal visibility, not only to improve lag recovery for more sources, but also to mitigate signal aliasing. We note that seasonal gaps may not affect lag recovery as much for sources with long variability timescales, but long baseline monitoring is required in such cases. 

There are complex interactions between the window function, emission-line lag and the relaxation time of the variability, which are further complicated by the redshift time dilation effects. As industrial-scale RM surveys target large AGN samples over wide redshift and luminosity distributions, it will be difficult to find a `one-size-fits-all' solution to survey design. Careful optimisation of target lists of sources to be monitored will likely deliver significant survey success rates. Future surveys should consider a multi-tiered observing strategy to ensure that the window function is optimised for a more diverse range of sources. Extending the wavelength range of spectrographs to the near-IR will be a critical requirement for joint emission-line lag analyses.

\section*{Acknowledgements}

We thank the anonymous referee for their comments that improved the paper. UM and AP are supported by the Australian Government Research Training Program (RTP) Scholarship. PM and ZY are supported in part by the United States National Science Foundation under Grant No. 161553 to PM. PM is grateful for support from the Radcliffe Institute for Advanced Study at Harvard University. PM also acknowledges support from the United States Department of Energy, Office of High Energy Physics under Award Number DE-SC-0011726. TMD and JC are supported by an Australian Research Council Laureate Fellowship (project number FL180100168). 

We acknowledge parts of this research were carried out on the traditional lands of the Ngunnawal and Ngambri peoples. We pay our respects to their elders past, present, and emerging.

This analysis used \texttt{NumPy} \citep{harris2020array}, \texttt{Astropy} \citep{astropy:2013, astropy:2018}, \texttt{SciPy} \citep{2020SciPy-NMeth}, and \texttt{thorsky} (\url{https://github.com/jrthorstensen/thorsky}). Plots were made using \texttt{matplotlib} \citep{Hunter:2007} and \texttt{seaborn} \citep{Waskom2021}. This work has made use of the SAO/NASA Astrophysics Data System Bibliographic Services.

%%%%%%%%%%%%%%%%%%%%%%%%%%%%%%%%%%%%%%%%%%%%%%%%%%
\section*{Data Availability}

The simulation data will be made available upon reasonable request to the corresponding author. 
 
% The inclusion of a Data Availability Statement is a requirement for articles published in MNRAS. Data Availability Statements provide a standardised format for readers to understand the availability of data underlying the research results described in the article. The statement may refer to original data generated in the course of the study or to third-party data analysed in the article. The statement should describe and provide means of access, where possible, by linking to the data or providing the required accession numbers for the relevant databases or DOIs.

%%%%%%%%%%%%%%%%%%%%%%%%%%%%%%%%%%%%%%%%%%%%%%%%%%

%%%%%%%%%%%%%%%%%%%% REFERENCES %%%%%%%%%%%%%%%%%%

% The best way to enter references is to use BibTeX:

\bibliographystyle{mnras}
\bibliography{refs} % if your bibtex file is called example.bib

%%%%%%%%%%%%%%%%%%%%%%%%%%%%%%%%%%%%%%%%%%%%%%%%%%

%%%%%%%%%%%%%%%%% APPENDICES %%%%%%%%%%%%%%%%%%%%%

% \appendix

% \section{Some extra material}

% If you want to present additional material which would interrupt the flow of the main paper,
% it can be placed in an Appendix which appears after the list of references.

%%%%%%%%%%%%%%%%%%%%%%%%%%%%%%%%%%%%%%%%%%%%%%%%%%

% Don't change these lines
\bsp	% typesetting comment

\label{lastpage}
\end{document}